\documentclass[usenatbib]{aa}

\usepackage{graphicx}
\usepackage[export]{adjustbox}
\usepackage{amsmath}
\usepackage{siunitx}
\usepackage{txfonts}
\usepackage{natbib}
\bibpunct{(}{)}{;}{a}{}{,}
\usepackage{float}
\usepackage{calc}
\usepackage{textcomp}
\usepackage{placeins}
\usepackage{color}
\usepackage{rotating} 
\usepackage{lscape}
\usepackage{url}
\usepackage{hyperref}
\usepackage{multicol}
\usepackage{subfig}
\usepackage[all]{draftcopy}
\usepackage{longtable}
\usepackage{array}
\usepackage{threeparttable}
\usepackage{lscape}
\usepackage{wasysym}
\usepackage{soul} 
\DeclareTextSymbol{\degre}{OT1}{23}
\newcounter{savedfootnote}

\usepackage[usenames,dvipsnames]{xcolor}
\usepackage{xspace}
\usepackage{xargs}
\usepackage{lineno}
\usepackage{amsbsy,amsmath}
\usepackage[utf8]{inputenc}
\usepackage{cleveref}

\newcommand{\galapagos}{{\sc galapagos}\xspace}
\newcommand{\galapagosII}{{\sc galapagos-2}\xspace}
\newcommand{\galfit}{{\sc galfit}\xspace}

\newcommand{\galfitm}{{\sc galfitm}\xspace}
\newcommand{\megamorph}{{MegaMorph}\xspace}
\newcommand{\sersic}{S\'ersic\xspace}
\newcommand{\sex}{{\scshape SExtractor}\xspace}
\newcommand{\psfex}{{\scshape PSFEx}\xspace}
\newcommand{\montage}{{\scshape montage}\xspace}

\newcommandx{\N}[2][1= ,2= ]{$\mathcal{N}^{#1}_{#2}$\xspace}

\newcommand{\Alpha}{A}

\begin{document}
   \title{Multi-wavelength structure analysis of local cluster galaxies.\\
   The WINGS project.}
\author{ A. Psychogyios\inst{\ref{inst1},\ref{inst2}} 
  \and  M. Vika\inst{\ref{inst2}} \and  V. Charmandaris\inst{\ref{inst1},\ref{inst2},\ref{inst3}} 
  \and  S. Bamford \inst{\ref{inst4}} 
  \and  G. Fasano \inst{\ref{inst5}}  
  \and  B. H{\"a}u{\ss}ler\inst{\ref{inst6}} 
  \and  A. Moretti \inst{\ref{inst5}}  
  \and  B. Poggianti \inst{\ref{inst5}} 
  \and  B. Vulcani \inst{\ref{inst7},\ref{inst5}} 
}
 	
	\institute{Department of Physics,  University of Crete, Herakleio, Greece\\\email{alpsych@physics.uoc.gr}\label{inst1}
	\and IAASARS, National Observatory of Athens, GR-15236, Penteli, Greece \label{inst2}
	\and Institute of Astrophysics, Foundation for Research and Technology-Hellas, GR-71110 Heraklion, Greece \label{inst3}
	\and School of Physics \& Astronomy, The University of Nottingham, University Park, Nottingham, NG7 2RD, UK \label{inst4}
	\and INAF, Osservatorio Astronomico di Padova, Vicolo Osservatorio 5, 35122 Padova, Italy \label{inst5}
	\and European Southern Observatory, Alonso de Cordova 3107, Vitacura, Casilla 19001, Santiago, Chile \label{inst6}
	\and School of Physics, University of Melbourne, VIC 3010, Australia \label{inst7}
	}


 
\abstract{ 
We present a multi-wavelength  analysis of the galaxies in nine clusters selected from the WINGS dataset, 
examining how galaxy structure varies as a function of wavelength and environment using the state of the art software \galapagosII. 
We simultaneously fit single-\sersic functions on three optical (u, B and V) and two near-infrared (J and K) bands thus creating a wavelength-dependent model of each galaxy. 
We measure the magnitudes, effective radius ($R_{e}$) the \sersic index ($n$), axis ratio and position angle in each band.
The sample contains 790 cluster members (located close to the cluster center < $0.64 \times R_{200}$) and 254 non-member galaxies that we further separate based on their morphology into ellipticals, lenticulars and spirals. 

We find that the \sersic index of all galaxies inside clusters remains nearly constant with wavelength while $R_{e}$ decreases as wavelength increases for all morphological types.
We do not observe a significant variation on $n$ and $R_{e}$ as a function of projected local density and distance from the clusters center.
Comparing the $n$ and $R_{e}$ of bright cluster galaxies with a subsample of non-member galaxies we find that bright cluster galaxies are more concentrated (display high $n$ values) and are more compact (low $R_{e}$).
Moreover, the light profile ($\mathcal{N}$) and size ($\mathcal{R}$) of bright cluster galaxies does not change as a function of wavelength in the same manner as non-member galaxies.  
}
    
   \authorrunning{Psychogyios et al. 2018}
   \titlerunning{Multi-wavelength structure analysis of local cluster galaxies.}

  \maketitle

\section{\label{intro}Introduction}
Having observed and statistically analysed a large fraction of the
galaxies in the local Universe it is widely accepted that only
few galaxies are found isolated in the field, while the majority
of them are preferentially located in denser environments such
as groups or clusters \citep{schmidt97,robotham11}.
Clusters of galaxies
are described as dense peaks in the galaxy distribution across the
sky, which were slowly created when the first massive dark matter
halos were decoupled from the nearly homogeneous expanding Universe,
and they contain hundreds up to thousands member galaxies.

Observationally, since all galaxies that belong to a cluster
are approximately at the same distance from us we can easily trace their
evolution as individual systems as well as how they are affected by the
properties of the cluster as a whole. 
Moreover, since the relative
distances between cluster galaxies is rather often comparable to
their size, processes such as ram pressure \citep{gunn_gott_72}, harassment (\citealp{moore96,moore98,moore99}), tidal forces \citep{byrdvaltonen90},
interaction/merging \citep{icke85,bekki98}, and
starvation/strangulation of star formation \citep{larson80} speed up galaxy evolution. 
Thus, the
study of local galaxy clusters is of paramount importance as they
enable us to easily quantify changes in the properties of their members
as a function of the baryonic density in the environment, and to
place constraints as a reference "zero-point" for comparison with
similar studies at higher redshifts.

Even though it was \citet{hubble_humason_31} who found
that ``the predominance of early-types is a conspicuous feature of clusters in general'', the systematic study of galaxy clusters
properties begun some decades later when improvements in observing
facilities enabled a more thorough study of their properties.
One of the first major findings 
was by \citet{butcher_oemler} who showed that the fraction of blue galaxies was higher
in clusters at $z$ $\gtrsim$ 0.4 than in nearby clusters.
This result was interpreted as the ageing of spiral galaxies after consuming their gas supply
and, therefore, diminishing their star formation rates.

\citet{dressler80} focused on the variation of galaxy properties
inside the clusters and found that early-type galaxies are
more abundant in the central part of galaxy clusters, a result
that is known today as morphology-density relation. In addition, \citet{whitmore93}
examined the same sample of 55 rich
clusters of \citet{dressler80} and they showed that the distance
from the cluster center is the driver of the galaxy evolution in
clusters for all galaxy types.

Despite the extensive studies, the details regarding the key physical mechanisms responsible for the
changes in galaxy morphology in dense environments, and how quickly that changes with redshift remains an open issue.
\citet{fasano00} showed that clusters at 0.1 $\leq$ z $\leq$ 0.25 are
a factor of 2-3 less abundant in spiral galaxies compared to those at intermediate redshifts
while the fraction of S0 galaxies is increasing compared with
the results at intermediate redshifts 0.4 $\leq$ z $\leq$ 0.5 \citep{dressler97,smail97}.
Moreover, the number of early-type
galaxies (ellipticals and S0s) decreases up to z $\sim$ 1 (\citealt{vandokkum00}, \citealt{lubin02}).
In addition, \citet{postman05} and \citet{desai07} 
demonstrated that the decline of early
type galaxies when we move to higher redshifts, i.e. z $\sim$ 0.8 - 1, 
is the result of the decreasing proportion of S0 galaxies. 
Finally, \citet{cerulo17} found that the red sequence at z $\sim$ 1
is dominated by elliptical galaxies at all luminosities and stellar
masses while the red sequence 
of local galaxy clusters becomes dominated by S0 galaxies.
This last finding may imply that ellipticals and S0 galaxies
follow different evolutionary paths.

A powerful method to characterise galaxy types is to study
the structure based on quantification of the integrated light profiles.
For instance, one of the most common analytic function,
the \sersic function \citep{Sersic68}, quantifies the structure
of galaxies in a few parameters such magnitude, the effective
radius ($R_{e}$), \sersic index ($n$) and ellipticity.
However, the apparent structure of a galaxy may change as a function of the wavelength
since different physical processes dominate the emission
at various wavelengths, while the extinction by dust may obscure
the central regions of galaxies, greatly affecting the interpretation
of the global morphology (\citealt{MacArthur04,labarbera10b,kelvin12,Kim2013,vulcani14,kennedy15}; Mosenkov et al. in prep.).

Studying the structure of galaxies in multiple bands and examining how it varies as
a function of wavelength enable us to better investigate the physical process 
that shape the galaxies. \citet{labarbera10b} made the first attempt based 
on a sample of 5080 local early-type galaxies in optical and NIR wavelengths. 
They showed that the \sersic index does not vary considerably with
wavelength while the mean effective radius decreases significantly
with increasing wavelength from g to H.
\citet{kelvin12} divided the Galaxy and Mass Assembly \citep[GAMA,][]{driver09} 
galaxies into early- and late-types using optical-near-infrared colors and 
\sersic index and showed that there is an observed change in galaxy size 
as a function of wavelength, which is probably caused by dust attenuation and/or the inside
out growth of galaxies. 
\citet{vulcani14} investigated further the GAMA sample applying a multi-wavelength 
analysis (the same to the present study)
and showed that there is a substantial
increase in \sersic index, and a decrease in effective radius
across the same wavelength range. It was proposed that
metallicity gradients as well as dust attenuation were the main
reasons driving these trends.
\citet{kennedy15} expanded the \citet{vulcani14} study and
found that early-type galaxies
show little variation in their \sersic index with wavelength but
they are significantly smaller at longer wavelengths. Late-type galaxies
(low-$n$) display a substantial increase in \sersic index with
wavelength.

In this paper, we build on the previous GAMA studies, which were
focused on field galaxies, 
we use a well tested methodology of multi-wavelength structural
modeling and apply it on optical and near NIR imaging of galaxies selected from {\textit{WIde-field Nearby Galaxy-cluster Survey}} \citep[WINGS,][]{fasano06}.
Furthermore, we combine the derived structural measurements with physical cluster 
parameters in order to investigate these affect the observed galaxy morphology. 
Our goal here is to examine how the galaxy structural properties as a function of wavelength are affected by the local environment.

The paper is organised as follows. In \S~\ref{data} we present the WINGS survey and the observations we used.
In \S~\ref{analysis} we present our analysis and we show the final sample.
In \S~\ref{results} we show the dependence of the structural parameters of the galaxies on the
wavelength used as well as on their location in the clusters, and compare them with galaxies in the field.
In \S~\ref{discussion} we compare our results with previous works. 
Finally, in \S~\ref{conclusions} we summarise and present our conclusions.

\section{WINGS survey}\label{data}
The WINGS is a wide field multi-wavelength imaging and spectroscopic survey of 77 nearby galaxy clusters (36 in the northern hemisphere and 41 in the south).
All clusters are in the redshift range 0.04$<$z$<$0.07, have a high Galactic latitude ($\mid$b$\mid$$<$20 deg) and they have been X-ray selected.

WINGS is the largest survey of nearby galaxy clusters that combines both optical and NIR imaging together with spectroscopic observations.
It provides a robust and homogeneous observational dataset suitable for the goals of this study.
Previous studies of WINGS provide catalogues with derived measurements such as morphology, local densities, distances from the brightest galaxy of the cluster, stellar masses, star formation histories, redshifts and cluster memberships (see \citet{moretti14} for more details on the data). 

The initial WINGS observations, in B- and V-bands (WINGS-OPT), consist of $\sim$550.000 galaxies. 
The optical images were taken using the wide field cameras on either the 2.5 m Isaac Newton Telescope (WFC@INT) or the MPG/ESO-2.2 m telescope (WFI@ESO). 
The pixel scales of WFC and WFI instruments are 0.332\arcsec and 0.238\arcsec respectively and they cover a Field of View (FoV) of 34$'$ x 34$'$. 
Additional B- and V-bands images were obtained from the wide-field imager OmegaCAM ($\sim$1 degree) of the VLT Survey Telescope (VST) on Paranal. 

Near infrared observations of the WINGS survey (WINGS-NIR) were performed with WFCAM@UKIRT and they contain $\sim$500.000 galaxies.
The NIR imagery has a pixel scale of 0.2\arcsec and a corresponding FoV of 0.79 degrees. 
A broad U-band imaging of a subsample of the WINGS clusters, has also been obtained with wide-field cameras at different telescopes (INT, LBT, Bok, see \citet{omizzolo14}). 
The pixel scale of U-band images is 0.21\arcsec.

Furthermore, spectroscopic information is available for 48 of the WINGS clusters with a high degree of completeness.
The spectroscopic observations were obtained with the AF2/WYFFOS multifiber spectrograph mounted on the 4.2 m William Herschel Telescope (WHT) and the 2dF multifiber spectrograph of the 3.9m Anglo Australian Telescope (AAT) (\citet{cava09}).
\citet{cava09} calculated the redshift of individual galaxies based on a semi-analytical method who identifies the emission lines of the spectrum.

\citet{vulcani12a} computed the projected local densities 
for the clusters, $\Sigma_{N}$, which is commonly defined as the number of neighbors ($N_{n}$) 
of each galaxy per $Mpc^{2}$.
They identified the 10 nearest neighbors of each cluster galaxy with $M_{V}$ $\leq$ $-19.5$ 
and calculated $\Sigma_{10}$=10$/\Alpha_{10}$, using a circle to define the area including these neighbors, where $\Alpha_{10}$=$\pi(R_{10})^2$ (Mpc) and $R_{10}$ (Mpc) is the radius (in Mpc) of the smallest circle centered on the galaxy.
\citet{fritz07,fritz11} calculated the integrated stellar masses for a subsample of 5229 WINGS galaxies via spectro-photometric modelling. 
They used the fiber spectra and the fiber/total magnitudes to derive the stellar masses, luminosity weighted and mass weighted ages, average star formation rates as well as the fiber and total magnitudes pertaining to the best fit model.
The morphological classification is based on
\citep[MORPHOT,][]{fasano12}.
MORPHOT produces two different morphological estimates based on a semi-analytical maximum likelihood technique and an neural network machine.
For more details about the WINGS survey see  also {\citet{varela09, valentinuzzi09, poggianti09, vulcani11a, vulcani11b, vulcani13, donofrio14}



For the purpose of this study we use 9 WINGS clusters with available imaging in at least four bands as well as spectroscopic redshift measurements.
Table \ref{properties} lists the available images for each cluster, the cluster name, coordinates, redshift, the available images and the number of member and non-member galaxies of each galaxy cluster that have been successfully fitted by our analysis.

\begin{table*} 
\caption{}
\label{properties}
\centering
\begin{tabular}{*{7}{c}} 
\hline\hline
\smallskip
Cluster & $\alpha$  & $\delta$ & $z$ & No. of cluster & No. of non-member & No. of cluster \\
            &  J2000     & J2000    &        & galaxies         & galaxies                   & galaxies ($M_{V}$ < -19.27 mag) \\
(1)&(2)&(3)&(4)&(5)&(6)&(7)\\
\hline
 A0119  & 00 56 21 & -01 15 & 0.0442  & 144 & 33 & 72\\
 A0500  & 04 38 52 & -22 06 & 0.0670 &  80 & 33 & 39\\
 A1291  & 11 32 21 & 55 58 & 0.0527 &  12 & 20 & 4\\
 A1631a & 12 52 52 & -15 24 & 0.0466 & 132 & 28 & 43\\
 A1983 & 14 52 59 & 16 42 & 0.0444 &  19 & 15 & 8\\
 MKW3s & 15 21 52 & 07 42 & 0.0453 & 25 & 10 & 13\\
 A2382 & 21 51 55 & -15 42 & 0.0644 & 162 & 55 & 69\\
 A2399 & 21 57 13 & -07 50 & 0.0582 & 112 & 45 & 40\\
 A2457 & 22 35 41 & 01 29 & 0.0591 & 104 & 15 & 50\\
\hline
\end{tabular}
\tablefoot{Columns: (1) Galaxy cluster name. (2) Right ascension. (3) Declination. (4) Redshift. (5) number of member galaxies in each galaxy cluster (no absolute magnitude cut). (6) number of non-member galaxies (with $z$ < 0.15 and  $M_{V}$ < -19.27 mag) in each galaxy cluster. 7) number of member galaxies in each galaxy cluster (with $M_{V}$ < -19.27 mag).
}    
\end{table*}

\section{Analysis}\label{analysis}

Given the size and the complexity of the datasets used we present in detail the necessary steps taken for the analysis.

\subsection{Rescaling the images}
The multi-wavelength fitting process performed in this study requires all images to be in the same resolution and to cover the same area of the sky.
To perform this, we use the \montage software \citep{berriman08}. 
First, all images are converted to the same pixel scale using as a reference the pixel scale of the K-band and then they are cropped to the size of the smallest image i.e. V-band image.  
We then correct for flux conservation by multiplying each optical image for the factor $(pix. scale_{K-band} / pix. scale_{opt. band})^2$. 
We convert all the optical images to counts multiplying each optical image with the given exposure time, as required by \galfitm.
Finally, we crop the images of each cluster in tiles of $2000\times2000$ pixels with an overlap of 500 pixels (49 tiles at each band) to make the subsequent processing by \galapagos \citep{barden12} (described in the next subsection) faster. 
We note that we calculate the zero point of each band based on the counts format.

\subsection{Creating different PSFs} 
Since the point spread function (PSF) varies 
depending on the physical location on the focal plane, we construct different PSFs along the x- and y-axis of the images for each band. 
We rely on the PSF Extractor code for this task \citep[see][\psfex]{bertin11} as well as the description of \sex \citep{bertin96}. 
\sex detects sources, and provides both photometric and shape information for all of them, which can be used by \psfex to identify a clean sample of sufficiently bright stars. We then \psfex to each image and construct 14x14 PSFs per filter and field, based on a regular grid as is usually done by \psfex. This means that the typical distance between PSFs is 700 pixels. 
Using the analysis plots provided by \psfex, we made sure that none of the PSF parameters such as FWHM or ellipticity changes significantly on such a scale. PSFs have also visually inspected for any abnormal features. \galapagos offers the possibility to use such a grid by choosing the closest PSF to each object for its fit. This way, it is assured that the optimum PSF is used at all times.


\subsection{Using the \galapagos software}\label{software}
\galapagos \citep{barden12} is an IDL wrapper of \galfit \citep{peng02} and \sex \citep{bertin96} that enables the automated detection and structural analysis of galaxies in a, typically large, image. 
\galapagos uses \sex to detect sources in the data, estimates a local sky background, cuts postage stamp images for all sources (smaller sections of the inputs images centered on individual sources, with dimensions based on the Kron radius of the source), prepares object masks, performs \sersic fitting taking into account light contamination from neighboring sources and compiles all objects in a final output catalogue.
In our study we make use of the updated version \galapagosII \citep{Haussler13} that utilizes a multi-wavelength version of \galfit, named \galfitm, developed by the \megamorph team \citep{bamford12,Haussler13,vika13,vika14,vika15}. 

We choose the V-band for detecting the sources for each cluster as it is deeper than the rest. 
The High Dynamic Range (HDR) mode of \galapagos \citep{barden12} detects the sources in two stages. First the so-called "cold run" detects (but not splitting up) all bright objects; the second, "hot", run detects fainter sources. 
In addition, we have changed the keywords in the first block of \galapagosII (cold and hot configuration files of HDR \sex source detection) in order to improve the splitting effect of cluster galaxies according to the image specifications of WINGS survey.
After combining the cold and hot catalogues, \galapagosII provides the data for an easy visual inspection. In this step it is easy to flag any misdetections (split up objects or fake detecting), which can be removed from any future part of the code. In this work, we examined all detections to ensure that all such misdetections are correctly flagged up.
For each detection \galapagosII cuts the science images into postage stamps. 
It also provides the sky background very efficiently using a flux growth method to estimate the local sky around an object \citep[see][]{barden12}. 
\galapagosII applies \galfitm in each postage stamp and models the light distribution of the central source.
It is worth noting that \galapagosII masks out distant and faint neighbours. Reasonably bright neighboring sources that can affect the model of the central source are fit simultaneously.

In addition, we make use of the \galapagosII feature that masks the deblending sources. Finally, we visually inspect the whole FoV of each band after the detection process to confirm that there is no artificial splitting of a single source.

After the fitting process, \galapagosII creates a catalogue with all calculated parameters from \sex and \galfitm.
For more details on the software performance see \citet{Haussler13}.


The fitting process is performed with \galfitm, which uses the
Levenberg-Marquardt (LM) algorithm to minimise the residual 
between a galaxy image and the PSF-convolved model by
modifying the free parameters.
In our setup we use the sigma images internally created by \galfitm.
For this study we use images of five different wavelengths in optical and NIR. 
As a first step, \galfitm fits a single, wavelength-dependent model to all images.
A second step allows for Bulge/Disk decomposition, but as this feature is not being used in this work, we are not discussing it here.
The model comprises a single-\sersic function, with a number of
parameters, e.g. center position (xc, yc), magnitude (m), effective radius ($R_{e}$), 
\sersic index ($n$), axial ratio ($Q$) and position angle ($PA$).

The magnitude input/starting value to \galfit is the ($m_{input}$) is the \sex MAG BEST
for the V-band image and typical offsets for the other images.
These offsets for each band are calculated to adjust (on average) magnitudes
measured on the V-band detection image to those for individual
bands. We allow full freedom in magnitudes, while we
allow the \sersic index and $R_{e}$ to vary with wavelength linearly
(as first order polynomials). 
We have also explored a quadratic variation of the \sersic index and $R_{e}$ with wavelength but we did not observe significant changes on the structural parameters. 
Since we have only 5 bands for the majority of clusters, we decided to allow the smallest variation of the structural parameters with wavelength. 
Moreover, allowing a minimal degree of freedom we optimally use the advantages that \galfitm offers.

The galaxy center position, the position
angle and the axis ratio are chosen to be constant with
wavelength.

There are multiple advantages of using simultaneously multiple
images for modeling the light distribution of a galaxy, such
as increasing the overall signal-to-noise and using the color difference
between the components to support decomposition. By constraining the variation freedom for some parameters, it reduces
the statistically uncertainty of the effective radius, \sersic
index and magnitude \citep{Haussler13}.


\subsection{Sample Selection}\label{sec:sample}

We validate the \galapagosII catalogue in order to select only the
objects that have been successfully modeled. In particular, each
source with one or more parameters lying on (or very close to)
a fitting constraint has been discarded.

We keep in our analysis only galaxies that satisfy the following criteria, as recommended by \citet{Haussler13} and \citet{vika13}.

\begin{itemize}
   \item $0 < m < 40$ at all wavelengths, where m is the total apparent output magnitude in each band.
   \item $m_{input}$  - 5 $<$ $m$ $<$ $m_{input}$ + 5, where $m_{input}$ is the starting value of the magnitude in each band.
   These are derived by scaling an average galaxy SED by the MAG BEST value measured by \sex during the object detection. 
During the fit, we allow a generous $\pm 5$ magnitudes variation from these starting values.
   \item $0.201 < n < 7.90$, since fits with values outside these ranges are rarely meaningful and very close to the constraints used during the fit. 
   \item $0.301 < R_{e} < 399.0$ pixels, which maintains values in a physically meaningful range and prevents the code from fitting very small sizes, where, due to oversampling issues, the fitting iterations become very slow.
   \item $0.1 < Q \leq 1.0$, where $Q$ is the axis ratio, to ensure the fit value is physically meaningful.
\end{itemize}

Finally, we clarify to the reader that the total number of galaxies excluded due to \galapagosII constraints 
are 13 member galaxies and 9 non-member galaxies (2$\%$ of our sample, thus it does not affect the uncertainty on the fitted parameters).
Equally we can argue that all our constraints are conservatively chosen, so wide that they only help to avoid unphysical areas of parameter space.

We focus on  galaxies with secure structural parameters, redshift measurement \citep{cava09} and morphological classification \citep{fasano12}. 
In Fig. \ref{color_mass_histograms} we present the B-V color and total stellar mass distributions for the galaxies inside the 9 clusters.

\begin{figure}\resizebox{\hsize}{!}{\includegraphics{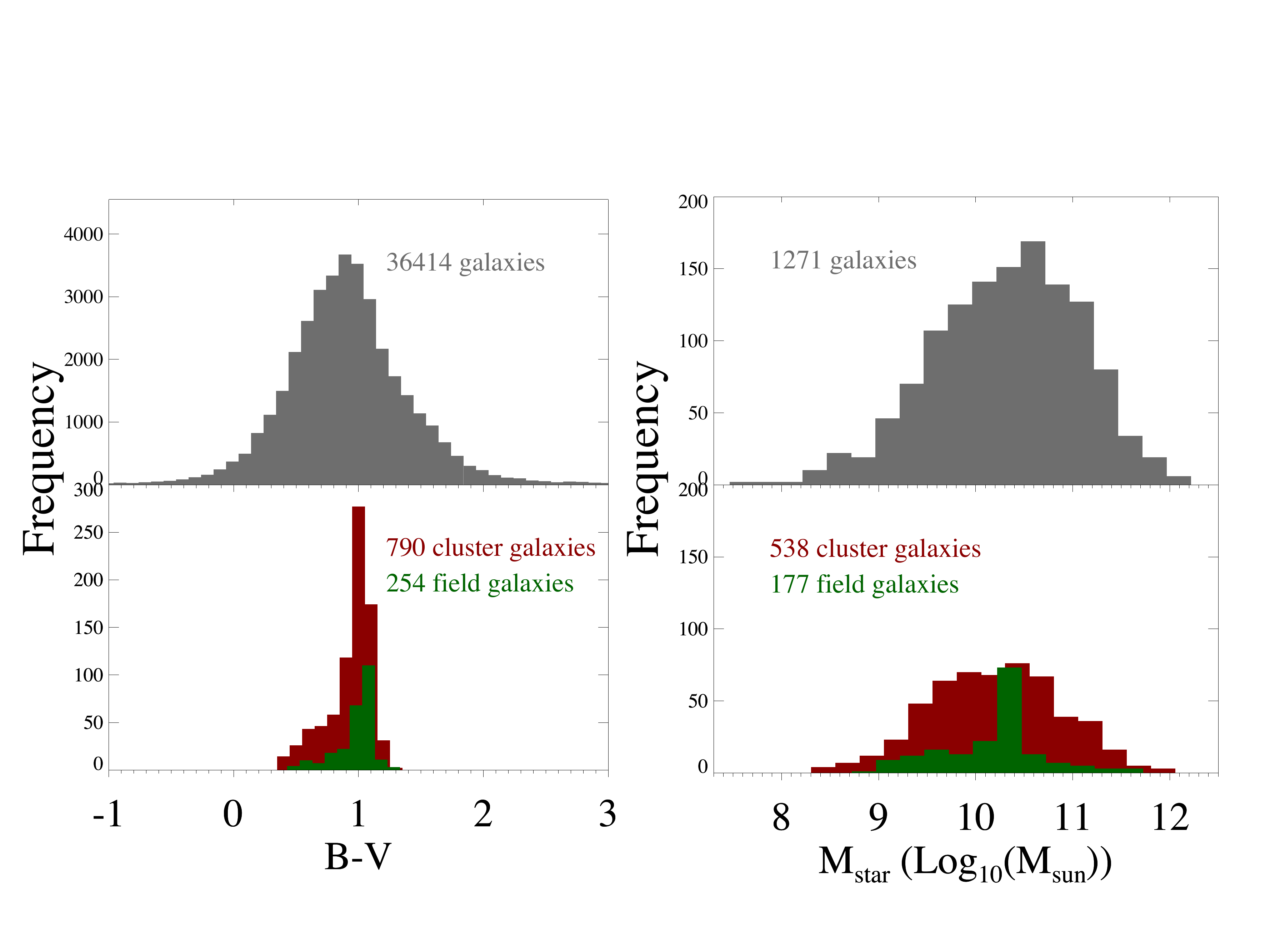}}
\caption{Left panel: B-V color and total stellar mass distributions for the galaxies inside the 9 clusters. On the upper panel we show the distribution for all WINGS galaxies of the 9 clusters (gray color) with available B-V colors from \citet{varela09} and at the bottom we show the histogram of the galaxies of this study (red for cluster galaxies and green for non-member galaxies). 
Right panel: Same as on the left, but for the stellar mass. Galaxies of this study have secure structural parameters (available redshift measurements and available morphological classifications).
The fact that the total number galaxies of this study (790) is much smaller than the 36414 galaxies of \citet{varela09} is due to the lack of spectroscopic data and morphological classification values. The lack of spectroscopic measurements is also responsible for the small number of galaxies (1271) with available stellar mass estimates. Our sample includes the majority (more than 95$\%$) of galaxies (both cluster members and non-members) that have spectroscopic measurements.}
\label{color_mass_histograms}
\end{figure}
In addition, in order to be consistent with geometrical biases, we limit our analysis at the minimum radius imaged for all clusters (0.64 $\times R_{200}$).

Our galaxy sample consists of 1044 galaxies, 790 of which are spectroscopically confirmed members.
The remaining galaxies (254) will constitute our non-members sample. 
These might be just field galaxies close to the clusters, or they could be indeed galaxies in small groups that will be soon falling into the clusters. 
The separation of cluster member and non-member galaxies was published in \citep{cava09} and \citep{moretti17}. We use the published results.
The absolute V magnitude ($M_{V}$) and the redshift range of cluster galaxies and non-member galaxies are presented in Fig. \ref{sample_volume_limited_sample_V}.
Table \ref{morphological_classes} tabulates the number of cluster and non-member galaxies in each cluster for each morphological class.

\begin{figure}[h] \resizebox{\hsize}{!}{\includegraphics{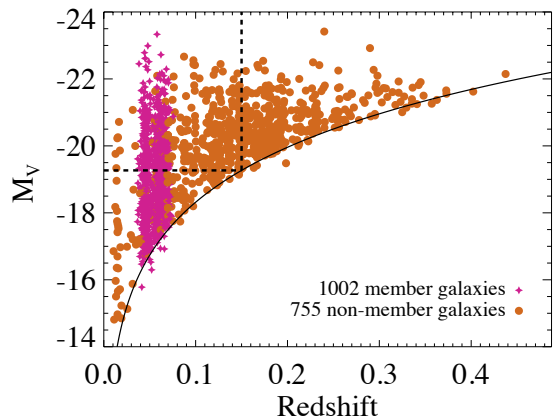}} 
\caption{Absolute V-band magnitude versus redshift for the detected galaxies (cluster and non-members). 
Purple red stars indicate the member galaxies while the brown circles are the non-member galaxies.
The curved line indicates the spectroscopic apparent magnitude limit of WINGS, V$\sim$ 20.0. 
The dashed vertical line represents the redshift cut at 0.15 while the horizontal dashed line indicates the $M_{V}$=-19.27.
}
\label{sample_volume_limited_sample_V}
\end{figure}

In order to study the difference between cluster galaxies and non-member galaxies we have created a secondary sample by applying the following selection criteria: At redshift z $\leq 0.15$ that corresponds to an $M_{V} < -19.27 mag$. 
The redshift cut (z $\leq 0.15$) helps to avoid galaxy evolution effects (Fig \ref{sample_volume_limited_sample_V}) but at the same time allows for sufficient number of galaxies to be included in the non-member sample. 
These criteria select 338 cluster galaxies and 254 non-member galaxies.
This second sample is only used in Section \ref{clusters_field}.

Based on the \citet{fasano06} classification (also \citet{calvi13} used this), we define as elliptical galaxies (E) those who have Hubble type (T) between -5.00 and -4.25, lenticular galaxies (S0) those with T $\ge -4.25$ and T $\le 0$, early-type spirals those with T > 0 and T $\le 4.00$ and late-type spirals the remaining galaxies (T $> 4.00$). 

We remind the reader that our sample is a magnitude limited sample ($m_{V}$ $\sim$ 20mag). 
We use all cluster galaxies (independent of their luminosity) to study their properties inside local clusters. 
In order to check if we are consistent with the completeness in luminosity and stellar mass we the do the following: 
We measure the ratio between the maximum and the minimum distance of the clusters (301 Mpc / 195 Mpc) $\sim$ 1.54.). The differrence in flux for those extreme clusters is equal to 2.38 which is interpreted in difference of 0.94 in magnitude. Equivalent will be the ratio in the luminosity limit. As for the stellar mass, we use the formula of \citet{bell03} ; 
M($M_{\astrosun}$) $\sim$ 0.95 $\pm$ 0.03 $\times$ $L_{K_{\astrosun}}$ 
and we find that the change in mass function sampling will also be a factor of $\sim$2.38 (depending of course on the B-R color corrections). 
However, the variation of structural parameters as a function of wavelength do not change if we use bright cluster galaxies (with $M_{V}$ < -19.27mag). 
For instance, the average $n$ values become larger while the average $R_{e}$ decreases.

At this point, we should mention that since four out of nine galaxy clusters do not have available J-band images we had to ensure that it did not introduce biases in our results. 
For that reason, we run our code on the galaxies in the clusters that have five available bands after removing the J-band, and we found no difference in the observed trends of $n$ and $R_{e}$.

Finally, for those galaxies that lack u- or J-band measurements, we estimate the u- and J-band \sersic index and effective radius from the results of B-, V-, K-band
taking into consideration that the structural parameters are varying linearly as a function of wavelength.

\begin{table}[!h]
\caption{}
\label{morphological_classes}
\centering
\begin{tabular}{*{5}{c}} 
\hline\hline
\smallskip
Cluster & Ellipticals  & S0s & Early-type  & Late-type  \\
  &   &  & spirals &  spirals \\
  \hline
A0119    &  50 (28) [8]   &  82 (37) [11]   & 12  (7) [13]   &    - (-) [1] \\
 A0500   &  23 (15) [9]   &  38 (15) [8]     & 14 (5) [12]    &  5 (4) [4] \\
 A1291   &    2 (-) [4]   &   7 (1) [9]      &   3 (3) [7]      &    - (-) [-] \\
 A1631a &  22 (9) [2]   & 60 (23) [10]    &  39 (11) [11]   & 11 (-) [5] \\
 A1983   &    4 (1) [5]   &   9 (5) [6]      &    6 (2) [4]     &    - (-) [-] \\
 MKW3s &   7 (5) [2]    & 13 (6) [3]     &    4 (2) [4]     &   1 (-) [1] \\
 A2382   & 34 (16) [8]    & 55 (29) [20]    &  51 (18) [17]  & 22 (6) [10] \\
 A2399   & 18 (8) [6]    & 42 (15) [13]    &  39 (16) [14]  & 13 (1) [12] \\
 A2457   & 29 (15) [3]    & 54 (23) [5]      &  18 (12) [6]    &   3 (-) [1] \\
\hline
\hline
\end{tabular}
\tablefoot{The sample of cluster members galaxies, cluster members galaxies with $M_{V}$ < -19.27 mag (inside parenthesis) and non-member galaxies (inside brackets) of the 9 galaxy clusters. Columns: (1) Galaxy Cluster name. (2) Number of elliptical galaxies. (3) Number of lenticular (S0) galaxies. (4) Number of early-type spiral galaxies. (5) Number of late-type spiral galaxies. 
}    
\end{table}

\section{Results}\label{results}
Since the stellar mass distribution is not the same for different morphological galaxy types \citep{vulcani12a}, we examine the dependence of the structural parameters (\sersic index and $R_{e}$) of the cluster galaxies as a function of the stellar mass.
It is obvious that in Fig. \ref{sample_n_total_stellar_mass} there is a clear trend, especially for ellipticals and lenticulars, between \sersic index and stellar mass. More massive early-type galaxies (ellipticals and S0s) have larger \sersic index values, while spiral galaxies have \sersic index values around 0.5 -1.5 independent of their total stellar mass.
On the other hand, we present in Fig. \ref{sample_re_total_stellar_mass} the dependence of $R_{e}$ on the stellar mass for all morphological types. 
As expected more massive galaxies display higher $R_{e}$ values. The stellar masses in Fig. \ref{sample_n_total_stellar_mass} and  Fig. \ref{sample_re_total_stellar_mass} derived from \citep{fritz11}. Since, not all WINGS clusters have available spectroscopic measurements some galaxies lack stellar masses. 
The stellar masses, star formation histories, extinction values and average stellar ages of galaxies were derived by analysing the integrated spectra of galaxies by means of spectral synthesis techniques.

\begin{figure}[h] \resizebox{\hsize}{!}{\includegraphics{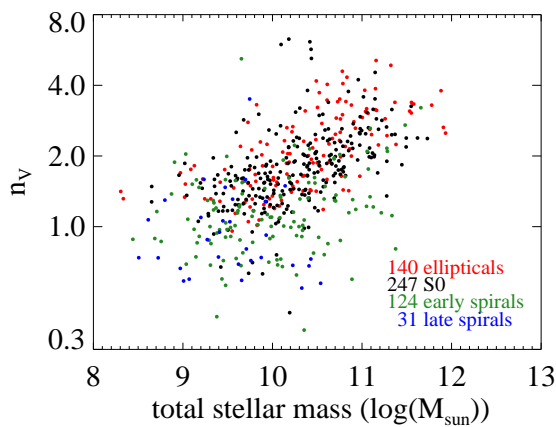}} 
\caption{The \sersic index in V-band as a function of the total stellar mass of the cluster galaxies sample up to $0.64 \times R_{BCG}$.  
The total number of the cluster members in this figure is lower because not all of them have available stellar mass measurements. 
}
\label{sample_n_total_stellar_mass}
\end{figure}

\begin{figure}[h] \resizebox{\hsize}{!}{\includegraphics{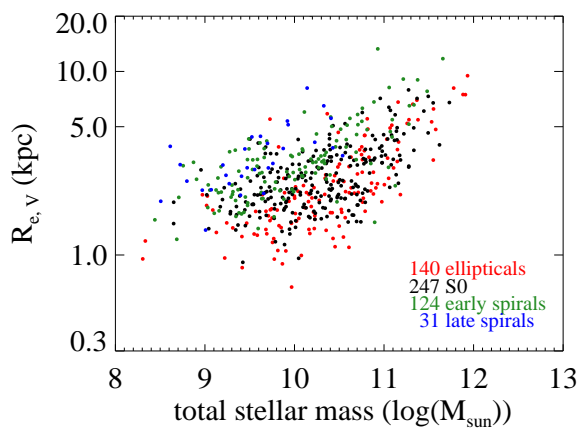}} 
\caption{The $R_{e}$ in V-band as a function of the total stellar mass of the cluster galaxies sample up to $0.64 \times R_{BCG}$. 
color coding and description is the same as in Fig. \ref{sample_n_total_stellar_mass}.
}
\label{sample_re_total_stellar_mass}
\end{figure}


\subsection{\textbf{The \sersic index of cluster members in the optical and NIR}}\label{sec:sersic}

In Fig. \ref{n_sample_histogram} we present a histogram of the
\sersic index for the confirmed cluster member galaxies in the
five bands. 
Elliptical galaxies have on average
the largest \sersic indices in all bands. 
Lenticulars (the most abundant galaxy type in our sample) display \sersic indices similar
to ellipticals but smaller on average. 
Early and late-type spirals have \sersic indices closer to 1 with the late-type spirals having
on average smaller values. 
We note that the scatter for the ellipticals and lenticulars is partially due to the nature of the sample that includes galaxies with different stellar masses as seen in Fig. \ref{sample_n_total_stellar_mass}.
\begin{figure}[h] \resizebox{\hsize}{!}{\includegraphics{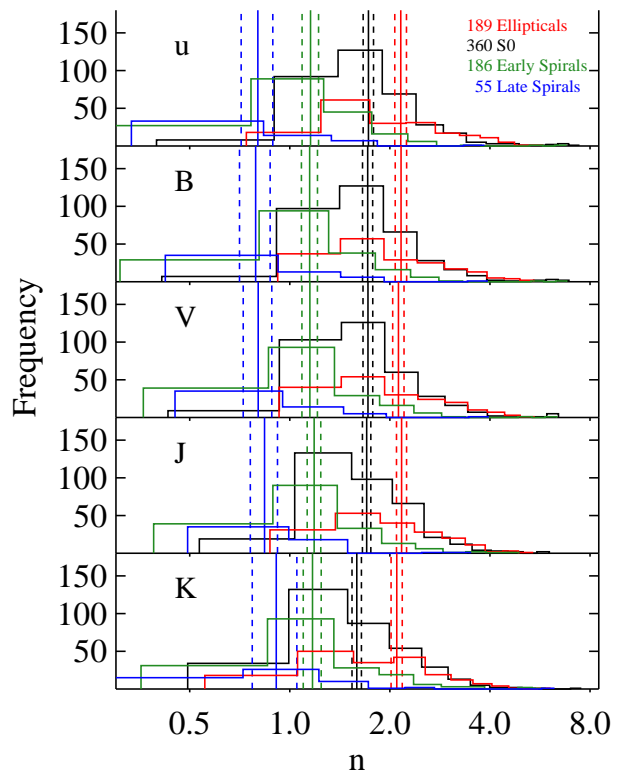}} 
\caption{Distribution of the \sersic index for the different bands for the cluster galaxies sample. 
Ellipticals, S0s, early and late-type spirals are in red, black, green and blue lines respectively. 
The vertical solid and dashed lines show the weighted median values and their corresponding uncertainty of each galaxy type calculated as $1.253 \sigma / \surd N$, where $\sigma$ is the standard deviation and N is the number of galaxies in the sample.
The legend indicates the number of galaxies per type.}
\label{n_sample_histogram}
\end{figure}

\begin{figure}[t!] \resizebox{\hsize}{!}{\includegraphics{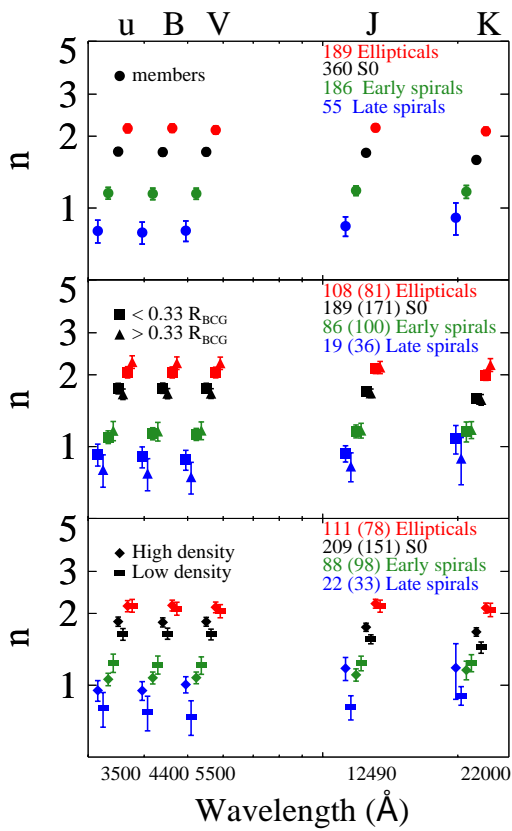} }
\caption{Weigthed median $n$ as a function of wavelength for different morphological types of galaxies. Color coding is the same as in Fig. \ref{n_sample_histogram}. 
The error bars show the uncertainty of the median (calculated as $1.253 \sigma / \surd N$, where $\sigma$ is the standard deviation and N is the number of galaxies in the sample).
The x-points are slightly shifted along the wavelength axis in order to make the errors readable. 
The numbers before the parenthesis show the total number of each galaxy type that are located in the inner region (middle panel) and inside high local density regions (bottom panel).
The numbers inside parentheses denote the total number of each galaxy type that are in the outer region (middle panel) and inside low local density regions (bottom panel).
}
\label{n_sample_multiplots_wt_med}
\end{figure}

In the top panel of Fig. \ref{n_sample_multiplots_wt_med} we present the weighted median $n$ of the cluster galaxies as a function of wavelength for the same four morphological types.
Error bars show the uncertainty of the median (calculated as $1.253 \sigma / \surd N$), where $\sigma$ is the standard deviation and N is the number of galaxies in the sample.
The spectroscopic weight for each sample galaxy analyzed is the inverse of the magnitude completeness factor $C(m)$ (see details in the Appendix B of \citet{fritz14}).
The weighted median value of our study is computed as follows:
We sort the structural parameters of a sample and their spectroscopic weights as well. 
We normalise the spectroscopic weights dividing with the total of the weights such as 

\begin{equation}
$$\sum_{i=1}^{n} w_{i, spec} = 1$$
\end{equation}

The weigthed median of the sample is the k-element parameter that satisfies the two conditions:

\begin{equation}
$$\sum_{i=1}^{k-1} w_{i, spec} \le1/2$$
\end{equation}

and 
\begin{equation}
$$\sum_{i=k+1}^{n} w_{i, spec} \le 1/2$$
\end{equation}

In the top panel of Fig. \ref{n_sample_multiplots_wt_med} we see that the weighted median $n$ of ellipticals and early-type spirals remains constant from the u-band to the K-band.
S0s have similar $n$ values from u- to J-band with slightly lower values in the K-band, while late-type spirals tend to display larger values in the K-band compared to the remaining bands.
We applied a Kolmogorov-Smirnov (K-S) test on the unbinned data of late-type spirals on the top panel of Fig. \ref{n_sample_multiplots_wt_med}. The corresponding probability of the K-S distribution between K- and the other bands (u, B, V and J) is (0.57, 0.42, 0.42, 0.88). 
From the previous KS test probabilities we cannot conclude if the Sersic index estimated in the K-band is different (or comes from the same sample) from the other bands.
We thus conclude that the median values of the \sersic index of late-type spirals remain constant from optical to NIR wavelengths.



In the middle panel of Fig. \ref{n_sample_multiplots_wt_med}, we examine how the
\sersic index changes as a function of the wavelength, but this time
after splitting the sample of the cluster member galaxies based on their
distance from the center of the corresponding cluster. Following the definition of \citet{cava09} the $R_{BCG}$  of a galaxy is the projected distance from the brightest cluster galaxy (BCG). 
The $R_{BCG}$ is normalized into $R_{200}$\footnote{$R_{200}$ is defined as the radius delimiting a sphere with interior mean density 200 times the critical density, approximately equal to the cluster virial radius. 
} units. 
Following the findings of \citet{fasano15}, we divide the area of the clusters into two
regions: the inner region, which extends out to $0.33 \times R_{200}$, and
the outer region, which is beyond this limit. We find that the inner region
contains 402 cluster member galaxies (108 ellipticals, 189 S0s,
86 early-type spirals and 19 late-type spirals) while the outer region consists of
388 galaxies (81 ellipticals, 171 S0s, 100 early-type spirals and 36 late-type
spirals).

It appears that the average $n$ of elliptical and early-type spirals remains the same independent of the location of the galaxies at the inner or outer regions of clusters and does not change as a function of wavelength.
However, lenticulars and late-type spirals located in the outer region of clusters have a slightly smaller average $n$  in all five bands compared to the galaxies in the inner region of the clusters.
Additionally, lenticulars located in the outer region of clusters show a drop of their average $n$ as we move from optical to NIR bands while late-type spirals located in the inner part of the clusters show an increase of their average $n$.


In the lower panel of Fig. \ref{n_sample_multiplots_wt_med} we split the sample based on the density of the environment where a cluster galaxy is located.
The first group found in regions with high projected local densities (log($\Sigma_{10}) > 1.45$) contains  430 cluster galaxies (111 ellipticals, 209 S0s, 88 early-type spirals and 22
late-type spirals) while the second, found in regions with low projected local densities (log($\Sigma_{10}) < 1.45$), contains 360 cluster galaxies (78 ellipticals, 151 S0s, 98 early-type spirals and 33 late-type spirals).
The values of the projected local density for each galaxy as well as the local density limit are based on the findings of \citet{fasano15}. 
We see that the projected local density is a cluster physical parameter that differentiate the average $n$ for all morphological types with the exception of the ellipticals.
S0s and late-type spirals have average $n$ larger in denser environments while early-type spirals show the opposite trend.
We applied K-S tests for all morphological types of galaxies to examine the results in the following manner: 
We compared all the elliptical galaxies (unbinned data) that were found in high-density regions against the ellipticals that reside in the low-density regions. We repeated the same approach for the lenticulars, early- and late-type spirals.
In particular, ellipticals do not change their average $n$ values as a function of wavelength (K-S test probability=0.05) in both high and low projected density regions.
The \sersic of lenticulars displays a decrease in J- and K-band that is stronger for galaxies found in low-density environment.   
The \sersic index of early-type spirals changes as a function of wavelength for regions with different projected local densities (K-S test probability=0.005) in both high and low projected density regions.
Late-type spirals found both in low and high density environments show a trend as a function of wavelength, where the average \sersic index is increasing as we move from u- to K-band but with an increase in the scatter as well.
The probabilities of a K-S test upon S0s and late-type spirals concerning low and high local densities are extremely low (< $10^{-5}$).

\subsection{\textbf{The effective radius of cluster member galaxies in the optical and near-IR}}\label{sec:re}

In Fig. \ref{re_sample_histogram} we present the distribution of the effective radius ($R_{e}$) for the different bands for our cluster member sample. 
The $R_{e}$ of each galaxy has been converted to physical units (kpc) using the radial velocity \citep{cava09} for each galaxy.
Spirals have the larger values of $R_{e}$ in all bands while the ellipticals and S0s have the smaller $R_{e}$ in all bands. 
The general trend for all types of galaxies is that the weighted median $R_{e}$ decreases from the optical to near-IR wavelengths.

In the top panel of Fig.  \ref{re_sample_multiplots_wt_med} we show the weighted median $R_{e}$
of the cluster member galaxies. 
All morphological types show a decrease of their average $R_{e}$ as a function of the wavelength with the elliptical and the early-type spiral to show the larger change.

\begin{figure}[t!] \resizebox{\hsize}{!}{\includegraphics{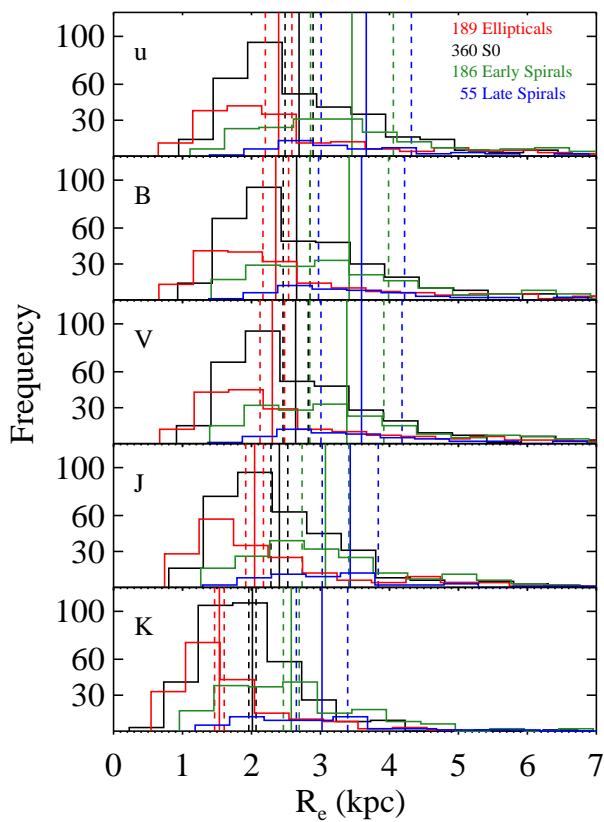}} 
\caption{Distribution of $R_{e}$ for the different bands for the cluster member galaxies.
The color coding and description are the same as in Fig. \ref{n_sample_histogram}.
}
\label{re_sample_histogram}
\end{figure}

\begin{figure}[t!] \resizebox{\hsize}{!}{\includegraphics{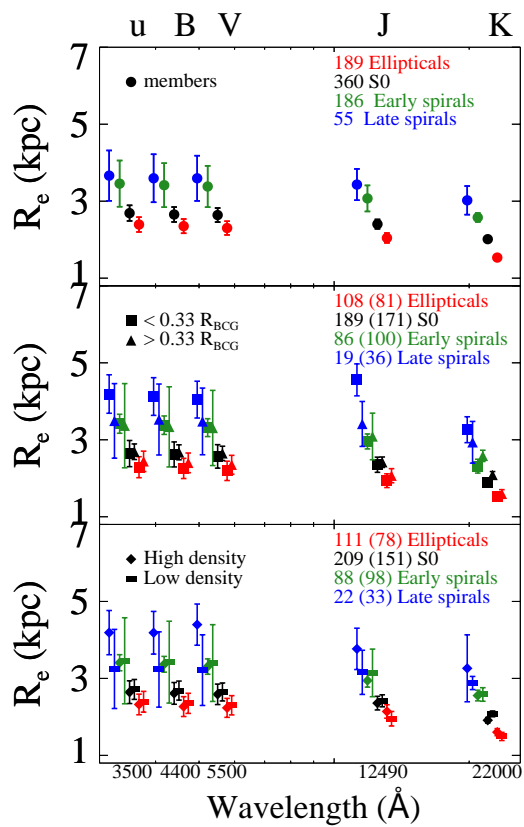} }
\caption{Weigthed median $R_{e}$ as a function of wavelength for different morphological types of galaxies.  
The color coding and description is the same as in Fig. \ref{n_sample_multiplots_wt_med}. 
}
\label{re_sample_multiplots_wt_med}
\end{figure}

In the middle panel of Fig. \ref{re_sample_multiplots_wt_med} we present how the effective radius changes as a function of wavelength for the galaxies located in the inner and outer region of the clusters. The sample separation is the same as in Section \ref{sec:sersic}.
We notice that elliptical, lenticular  and early-type spiral galaxies have the same average $R_{e}$ independent of their location in the cluster.
However, late-type spiral galaxies appear to have smaller average $R_{e}$ when they are at the outer area of the cluster, even though the difference is within the error bars.


In the bottom panel of Fig. \ref{re_sample_multiplots_wt_med} we examine how the $R_{e}$ of the cluster members changes in regions with low and high projected local densities. 
We see that for all galaxy types (except for the late-type spirals) there is no difference in the weighted median $R_{e}$ values in regions with high and low projected local densities.
The median value of late-type spirals $R_{e}$ is larger in higher local densities rather than in regions with lower local densities.
A KS test analysis reveals that the probability of low and high- density late type galaxies are drawn from the same sample is 0.004 (p-statistic equal to 1.00), which supports this result. However, this is not the case for the other galaxy types, since the KS p-statistics are equal to 0.70, 0.21, 0.69, (D values are equal to 0.40, 0.60, 0.40) for ellipticals, S0s and early type spirals, respectively.

\subsection{\textbf{Dependence on the distance from the BCG and the projected local density.}} \label{sec:fund_properties}
In this section we examine how $n$ and $R_{e}$ change as a function of the
projected distance from the cluster center ($R_{BCG}$) and the projected local density ($\Sigma_{10}$). 
In Fig. \ref{n_vs_rBCG_multi} we show the median values of $n$ in u-, V- and K-band
as a function of $R_{BCG}$.
We do not show the results of the B- and J-band as they are practically identical to those of V- and K-band respectively.
Elliptical, lenticular and early-type spiral galaxies have statistically constant $n$ values in all bands with late-type galaxies showing the smallest scatter.
In late-type spirals galaxies the average $n$ decreases up to $0.3\times R_{200}$ and remains constant within the errors further out. 

To investigate the robustness of the trend, we applied a linear least square fit to explore the dependence of $R_{BCG}$ as a function of the \sersic index for the unbinned data of late-type spirals that are located in [0.0 - 0.35] $\times$ $R_{200}$, and examined the uncertainty, due to the scatter of the points, to the intercept and slope coefficients. 
We find that for the u-band, the intercept and slope are: 1.5$\pm$0.18, and -2.54$\pm$0.66, while for the V-band they are 1.45$\pm$0.15 and -2.26$\pm$0.57. 
We perform a second linear least square fit to the unbinned data of late-type spirals located in [0.09 - 0.35] $\times$ $R_{200}$ excluding the only late-type galaxy found within [0.0 - 0.1] in order to  examine the trend with higher accuracy. 
We found that for the u-band the corresponding intercept and slope are 1.39$\pm$0.23, and -2.16$\pm$0.83, while for the V-band they are: 1.38$\pm$0.20 and -2.00$\pm$0.72. 
This suggests that the trend is consistent within 3$\sigma$ (u: 2.6 < 3; and V: 2.78 < 3) for galaxies located in [0.0 - 0.35] $\times$ $R_{200}$.
At largest distances from the cluster center the \sersic index for late-type galaxies remains nearly constant.

\begin{figure}[h] \resizebox{\hsize}{!}{\includegraphics{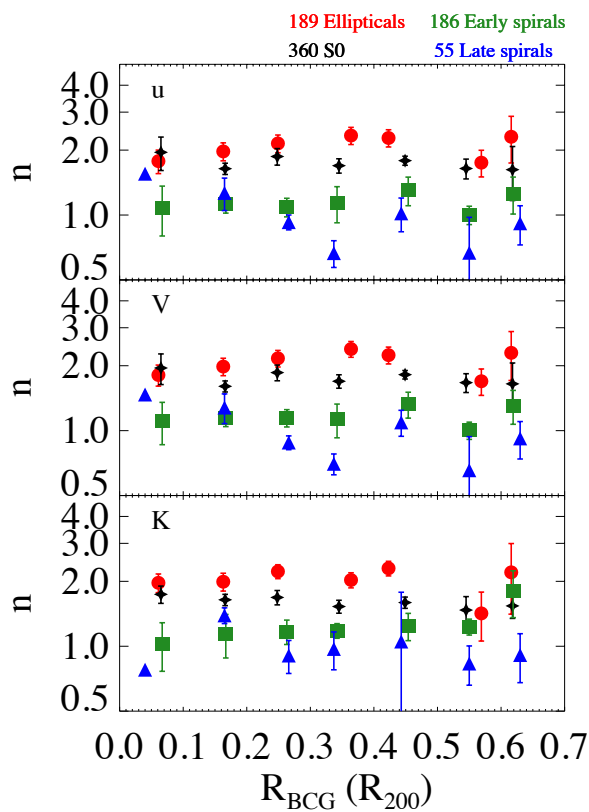}}
\caption{Weighted median values of $n$ in u-, V- and K-band as a function of $R_{BCG}$. 
The values of $R_{BCG}$ are normalized into $R_{200}$ units and the x-axis is divided into $R_{BCG}$ bins of $0.1 \times R_{200}$.
The error bars of each galaxy type in each bin are calculated as $1.253 \sigma / \surd N$, where $\sigma$ is the standard deviation and N is the number of galaxies of each galaxy type in each bin.
The x-points are slightly shifted in order to make the errors more readable and the color coding is the same as in Fig. \ref{n_sample_histogram}. 
Points with no error bars are those with one galaxy inside the corresponding bin.
Note that the number of galaxies per bins is not constant (see Table \ref{bins_1}).
}
\label{n_vs_rBCG_multi}
\end{figure}

\begin{table}[!h]
\caption{}
\label{bins_1}
\centering
\begin{tabular}{*{4}{c}} 
\hline\hline
\smallskip
Morphological class & minimum & maximum & mean   \\
  \hline
Ellipticals  & 7  & 37 & 27      \\
S0s  &  12 &  81 & 51      \\
Early-type spirals & 6  & 40 &  27     \\
Late-type spirals & 1 & 13 & 8      \\
\hline
\end{tabular}
\tablefoot{Minimum, maximum and mean values of each morphological class of galaxies inside bins of Fig. \ref{n_vs_rBCG_multi} and Fig. \ref{re_vs_rBCG_multi}.
}    
\end{table}

In Fig. \ref{re_vs_rBCG_multi} we examine the $R_{e}$ as a function of the distance from the BCG in optical and NIR. 
Elliptical and lenticular galaxies have fairly constant $R_{e}$ values in all bands. 
In early and late-type spiral galaxies the $R_{e}$ shows an increasing trend from the center out to [0.3 - 0.4] $\times$ $R_{200}$ in all bands, decreasing further out, even though the errors are large.
In order to investigate that trend, we examine again the coefficients of a linear least square fit for the unbinned data of early- and late- type spirals that are located in [0.0 - 0.35] `$\times$ $R_{200}$ and in [0.35 - 0.64] $\times$ $R_{200}$. 
We found that the early-type spirals show no correlation across $R_{BCG}$, with the following values for the intercept and slope: u-band: 3.30$\pm$0.56 and 1.40$\pm$2.34, V-band: 3.17$\pm$0.53 and 1.45$\pm$2.20, K-band: A=2.17$\pm$0.41 and 1.81$\pm$1.69. 
In contrast, late-type spirals show hints of an increase for $R_{e}$ up to 0.35 $\times$ $R_{200}$, since the intercept and slope are in the u- band: 2.30$\pm$1.07 and 6.73$\pm$3.98, while in the V-band: 2.18$\pm$1.02 and 6.74$\pm$3.78. 
Especially in K-band, the increase of $R_{e}$ is more significant with an intercept of 1.15$\pm$0.66 and slope 6.82$\pm$2.49.

\begin{figure}[h] \resizebox{\hsize}{!}{\includegraphics{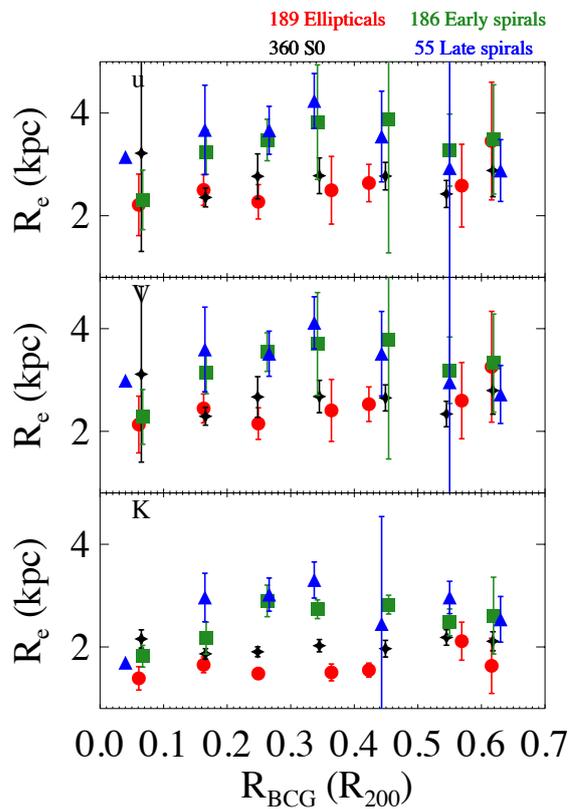}}
\caption{Weighted median values of $R_{e}$ in the u-, V- and K-band as a function of $R_{BCG}$. 
The color coding and description are the same as in Fig. \ref{n_vs_rBCG_multi}
Note that the number of galaxies per bins is not constant (see Table \ref{bins_1}).
}
\label{re_vs_rBCG_multi}
\end{figure}






In Fig. \ref{n_vs_LD_multi} we present the median values of $n$ in the u-, V- and K-band as a function of the projected local density.
Ellipticals, S0 and early-type spirals have no significant variation in their $n$ values, contrary to the late-type spirals where $n$ appears to 
increase in particular in the optical bands.
For the latter, we applied a spearman rank test on the unbinned data of late-type spirals. In optical bands (u,V) we do not see a clear trend. 
The rank correlation coefficient 
for u- and V-band is close to 0.
Regarding the K-band, (where we trace well the stellar mass of the galaxies, so the bulge is more prominent) we see that there is a weak trend (rank correlation coefficient is 0.3
and the two-sided significance of its deviation from zero (p statistic) is 0.025)
suggesting a slight increase of the \sersic index with projected local density.
\begin{figure}
\resizebox{\hsize}{!}	
{\includegraphics[width=\hsize]{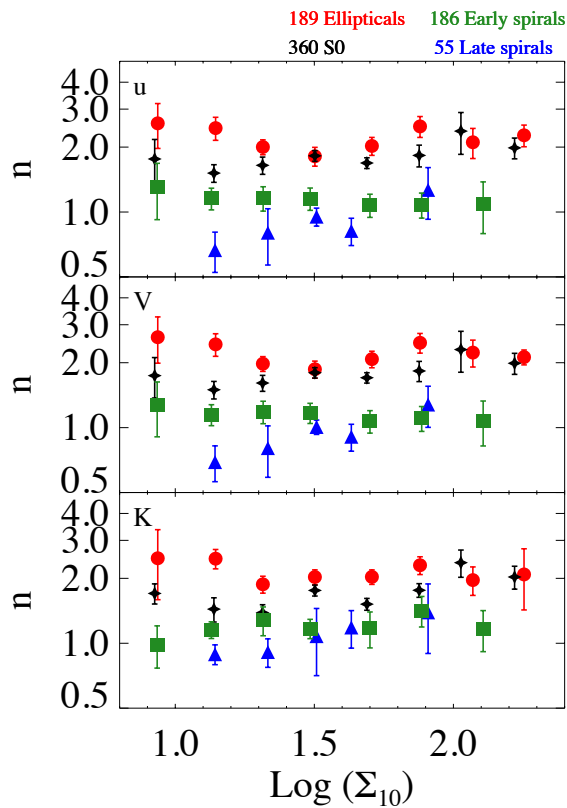}} 
\caption{Weighted median values of $n$ in the u-, V- and K-band as a function of the projected local density of the clusters. 
We divide the x-axis into projected local density bins of $0.2 \times$ log($\Sigma_{10}$).
The color coding and description are the same as in Fig. \ref{n_vs_rBCG_multi}
Note that the number of galaxies per bins is not constant (see Table \ref{bins_2}).
}
\label{n_vs_LD_multi}
\end{figure}

In Fig. \ref{re_vs_LD_multi} we show the median values of $R_{e}$ in the u-, V- and K- band as a function of the projected local density.
It is clear that the scatter of $R_{e}$ in all galaxy types is large. 
Even though, ellipticals and S0s do not change their size as a function of the local density in all wavelengths,
while 
late-type spiral galaxies tend to have larger $R_{e}$ in regions with very large local density values. 
We applied a spearman rank test on the unbinned data of early and late-type spirals. 
The rank correlation coefficient has low values in all bands and the significance of the relation between $R_{e}$ and local density is very low both for early and late-type spirals.
Furthermore, we investigate the effect of stellar mass upon the $n$ and $R_{e}$ versus projected local density. We divide the sample using a stellar mass cut of $log_{10}(M_{\star})$=10. 
We applied spearman tests upon the spiral galaxies in u-, V- and K-band. 
We observed a weak trend in the K-band for late-type spirals with high values of stellar mass. For that reason, we applied a Spearman test and we found that the Spearman coefficient is equal to 0.6 and the p-statistic is not very small (0.21). We conclude that the $R_{e}$ of high stellar mass late spirals and local density in the K-band do not show a correlation.

\begin{figure}
\resizebox{\hsize}{!}	
{\includegraphics[width=\hsize]{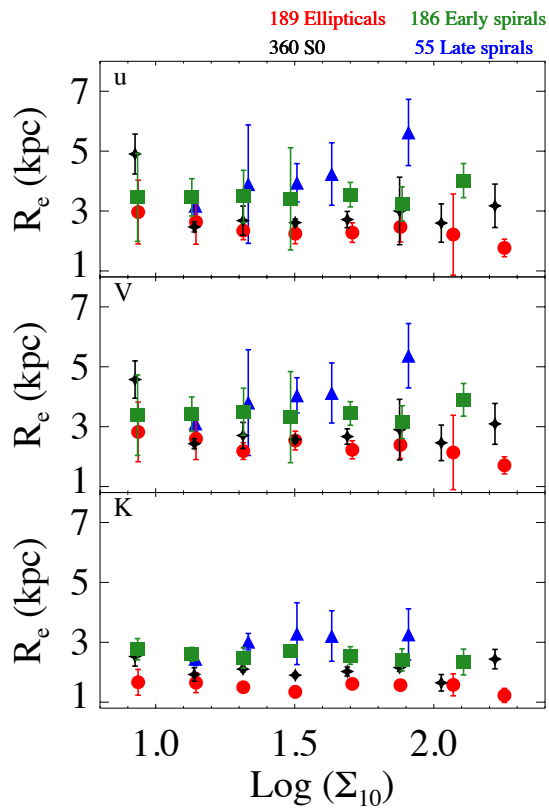}} 
\caption{Weighted median values of $R_{e}$ in u-, V- and K-band as a function of the projected local density of the clusters. 
We divide the x-axis into projected local density bins of $0.2 \times$ log($\Sigma_{10}$).
The color coding and description are the same as in Fig. \ref{n_vs_rBCG_multi}
Note that the number of galaxies per bins is not constant (see Table \ref{bins_2}).
}
\label{re_vs_LD_multi}
\end{figure}

\begin{table}[!h]
\caption{}
\label{bins_2}
\centering
\begin{tabular}{*{4}{c}} 
\hline\hline
\smallskip
Morphological class & minimum & maximum & mean   \\
  \hline
Ellipticals  & 4 & 46 & 24      \\
S0s  & 3 & 103 & 45      \\
Early-type spirals & 0 & 60 & 23      \\
Late-type spirals & 0 & 18 & 7      \\
\hline
\end{tabular}
\tablefoot{
In this table we report the minimum, the maximum and the mean number of galaxies used for calculating the values of \sersic index and $R_{e}$ in the bins of Fig. \ref{n_vs_LD_multi} and Fig. \ref{re_vs_LD_multi}. Each morphological class is reported separately.
}    
\end{table}

\subsection{\textbf{Structural properties of cluster members and non-member galaxies.}} \label{clusters_field}
\subsubsection{The \sersic index, $n$ } \label{sssfm}

As we discussed earlier (see \ref{sec:sample}) in addition  to the cluster member galaxies, our imagery also includes some  galaxies that are located close to the clusters.
We can use these non-member galaxies in order to contrast their morphological properties 
with the cluster member galaxies. 
With this comparison, we will be able to study galaxies evolving inside the gravitation potential of a cluster and galaxies that are located in the ``field''.

In Fig. \ref{members_field_sersic_plots} we present the weighted median values of \sersic index as a function of wavelength from optical to NIR for the cluster member and the non-member galaxy sample. 
We remind the reader that the cluster galaxies sample differentiate from the sample used in the previous plots (see Section \ref{sec:sample} for more details).
In general, the median $n$ of all galaxy types, in all the wavelengths, that live inside clusters are larger compared with galaxies located in the field. 
As we saw in Fig \ref{sample_n_total_stellar_mass}, ellipticals and S0s show a trend of increasing $n$ as a function of the stellar mass, thus ellipticals and S0s included in this secondary sample have higher average $n$ (and $R_{e}$) compared to the initial sample because this sample contains brighter (thus more massive) galaxies.

Non-member elliptical galaxies show the same trends with cluster ellipticals, that
$n$ stays constant as we move from optical to NIR
(the slope of the least square fit of cluster and non-member ellipticals are both negative and extremely low ($\sim 10^{-5}$).
S0 galaxies present different trends. 
The \sersic index of cluster S0s tends to decrease from the optical to the NIR (very low, positive least square slope) while the median $n$ values of non-member S0s increases slightly with wavelength (very low, negative least square slope).
Spiral galaxies (both cluster and non-member galaxies) tend to increase their median $n$ values as a function of wavelength (positive least square slopes)
but within the errors.
In general, we make KS tests among all bands, comparing two different bands each time and we found that the \sersic index KS test probabilities are $>$ 0.05 for all morphological types. We conclude that the \sersic index remains constant as a function of wavelength.
In Table \ref{change_wavel} we present the (\%) change of $n$ and $R_{e}$ and  from u- to K-band for cluster and non-member galaxies as presented in Fig. \ref{members_field_sersic_plots} and \ref{members_field_radius_plots}. 
The (\%) changes are calculated from the points (weighted median values, binned data) of Fig. \ref{members_field_sersic_plots} and \ref{members_field_radius_plots}. 
For example, the (\%) $R_{e}$ change of ellipticals (44\%) is calculated from the 100\% difference between the points in Fig. \ref{members_field_radius_plots} using the formula 
\begin{equation}
100 \frac{median(R_{e,K}) - median(R_{e,u})} {median(R_{e,u})}
\end{equation} 
The equivalent $R_{e}$ change error ($\pm$6\%) is calculated with the error propagation theory formula, which is equal to
\begin{equation}
100 \frac{1}{|median(R_{e,u})|} \sqrt{{\sigma(R_{e,K})}^2 + {(\frac{\sigma(R_{e,u})}{R_{e,u}} R_{e,K}})^2}
\end{equation}

\begin{figure}
\resizebox{\hsize}{!}
{\includegraphics{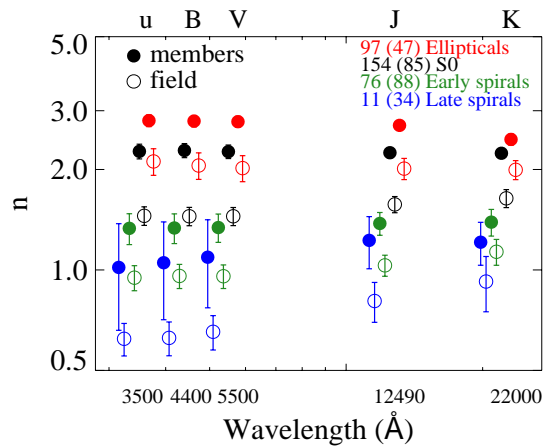} }
\caption{Median $n$ as a function of wavelength for cluster member and non-member galaxies. 
Asterisks indicate the cluster sample while filled crosses show the non-member sample. 
The color coding and description are the same as in Fig. \ref{n_sample_multiplots_wt_med}.
}
\label{members_field_sersic_plots}
\end{figure}

\begin{figure}
\resizebox{\hsize}{!}
{\includegraphics{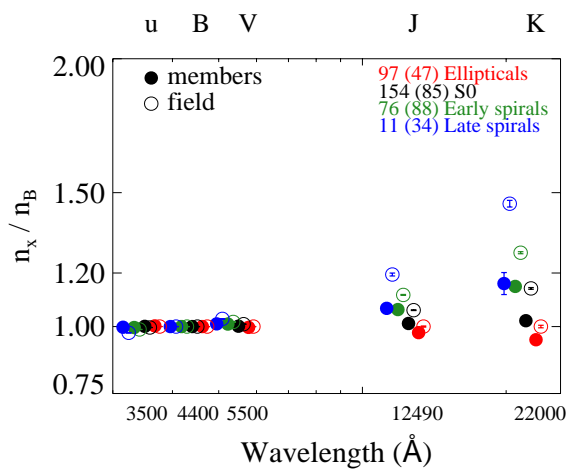}}
\caption{$\mathcal{N}$ for cluster member and non-member galaxies of different morphological types from optical to NIR. 
Asterisks indicate the median values of each band of member galaxies while filled crosses are the non-member galaxies.
The color coding and description are the same with Fig.\ref{n_sample_multiplots_wt_med}.
}
\label{sample_sersic_vulcani_fig9_v1}
\end{figure}


Figures \ref{n_sample_multiplots_wt_med} (top panel) and \ref{members_field_sersic_plots} illustrated how the weighted median $n$ changes across different wavelengths.
The sample of Fig. \ref{n_sample_multiplots_wt_med} contains the 790 cluster galaxies while Fig. \ref{members_field_sersic_plots} presents the sample of cluster galaxies with the non-member galaxies that have $M_{V} < -19.27 mag$.
Following \citet{vulcani14}, we can explore the variation of $n$ using the $\mathcal{N}_Y^X$=$n$(X)/$n$(Y) parameter, where $n$(X) and $n$(Y) are the \sersic index in X- and Y-band respectively.
Measuring the ratio of the \sersic index of different bands is a way to investigate the stellar populations and spatial structure of galaxies at different wavelengths. 
These ratios provide a simple but powerful parametrization of galaxy color gradients \citep{vulcani14,kennedy15}. 
Young and intermediate stars radiate mostly in the optical spectrum while older stellar populations emit in NIR. 
As a result, galaxies display a different structure due to the inherent distribution of different stellar populations and dust extinction inside the galaxies. 
Our multi-wavelength analysis provides a very good opportunity to compare values in different wavelengths due to the fact that the images are developed in a homogeneous process than fitting galaxies independently in each band.

\begin{figure} \resizebox{\hsize}{!}{\includegraphics{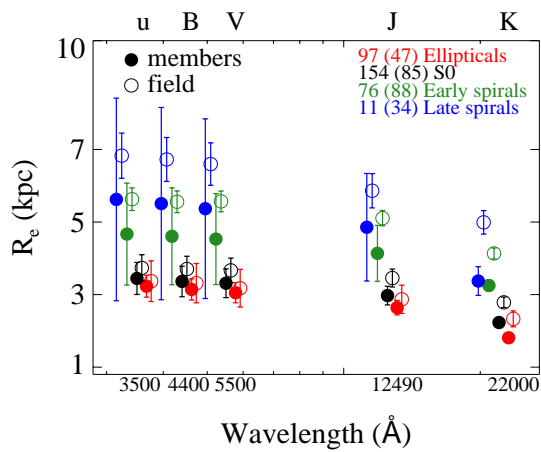} }
\caption{Median $R_{e}$ as a function of wavelength for cluster members and non-member galaxies. 
Asterisks indicate the cluster galaxies while filled crosses show the non-member galaxies. 
The color coding and description are the same as in Fig. \ref{members_field_sersic_plots}.
}
\label{members_field_radius_plots}
\end{figure}

\begin{figure} \resizebox{\hsize}{!}{\includegraphics{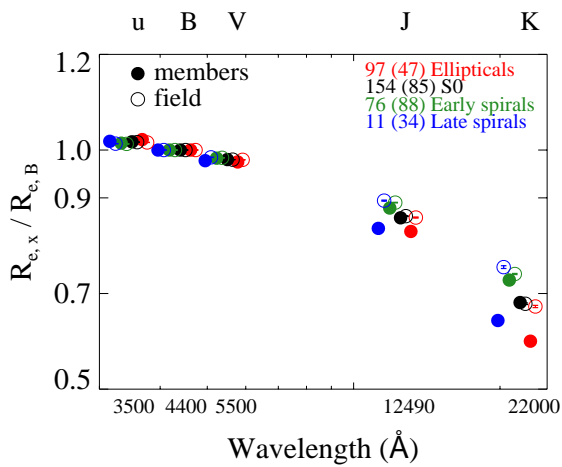}}
\caption{$\mathcal{R}$ for cluster members and non-member galaxies of different morphological types from optical to NIR.
The color coding and description are the same with Fig.\ref{sample_sersic_vulcani_fig9_v1}.}
\label{sample_radius_vulcani_fig16_v1}
\end{figure}

In Fig. \ref{sample_sersic_vulcani_fig9_v1} we use the B-band as the reference band and we calculate
the $\mathcal{N}$ of each band. 
We notice that cluster ellipticals have median $\mathcal{N}$ values less than 1 in the K-band while non-member ellipticals have median $\mathcal{N}$ values close to 1 in all bands.
Cluster lenticular galaxies do not change much the $\mathcal{N}$.
Both cluster and non-member spiral galaxies exhibit the same trend with increasing $\mathcal{N}$ values as the wavelength increases.
Finally, non-member spirals show larger variation compared to cluster members.

We investigate the influence of stellar mass in Fig. \ref{sample_sersic_vulcani_fig9_v1} by applying stellar mass cuts of $log_{10}$ ($M_{\star}$)=[9,10,11] in both cluster and non-member sample.
However, no changes in the variation of $\mathcal{N}$ for ellipticals and S0s, except for those with very high values of stellar mass  ($log_{10}$($M_{\star}$)>11), are visible.

\begin{table*}
\caption{}
\label{change_wavel}
\centering
\begin{tabular}{*{5}{c}} 
\hline\hline
\smallskip
Morphological & ($R_{e}$(K)/$R_{e}$(u) - 1)$\times$100 & ($R_{e}$(K)/$R_{e}$(u) - 1)$\times$100 & ($n$(K)/$n$(u) - 1)$\times$100 & ($n$(K)/$n$(u) - 1)$\times$100  \\
class & cluster galaxies  & non-member galaxies  & cluster galaxies  & non-member galaxies \\
  \hline
Ellipticals  & -44 $\pm$ 6  & -31 $\pm$13  & -12 $\pm$ 5  & -6 $\pm$ 10    \\
S0s  & -35 $\pm$ 9  & -25 $\pm$8  & -1 $\pm$ 6  & 13 $\pm$ 10   \\
Early-type spirals & -30 $\pm$ 21  & -27 $\pm$5  & 4 $\pm$ 15  & 20 $\pm$ 15       \\
Late-type spirals & -40 $\pm$ 30  & -27 $\pm$8  & 19 $\pm$ 45  & 49 $\pm$ 33        \\
\hline
\end{tabular}
\tablefoot{The (\%) change of $n$ and $R_{e}$ from u- to K-band for cluster and non-member galaxies as presented in Fig \ref{members_field_sersic_plots} and \ref{members_field_radius_plots}. 
(see section \ref{sssfm} for more details about the formulas we used to calculate the (\%) changes and their errors).}
\end{table*}

\subsubsection{The effective radius, $R_{e}$}

In Fig. \ref{members_field_radius_plots} we present the median values of $R_{e}$ as a function of wavelength from optical to NIR for the same sample of cluster members and non-member galaxies. 
We observe that non-member galaxies display larger $R_{e}$ compared to the member sample for all morphological types in all bands. 
Moreover, $R_{e}$ decreases from optical to NIR in a similar way for all galaxy types. 
(see Table \ref{change_wavel} for tabulated values).
The general conclusion of Fig. \ref{members_field_radius_plots} is that ellipticals have smaller $R_{e}$ than S0s, and S0s have smaller $R_{e}$ than early-type spirals. 
Cluster-member galaxies display lower $R_{e}$ compared to galaxies in the field. 
This statement holds only for S0, early-type spirals and late-type spirals. Ellipticals do not change their $R_{e}$ for different environments.
We applied a K-S test for each morphological class. For ellipticals we found that the KS \textbf{p-statistic} is equal to 0.70 and the \textbf{D value} is 0.40. For the remaining morphological classes, the \textbf{p-statistics} are less that 0.04 and the \textbf{D values} are above 0.80.
These results do not change if we include galaxies with higher stellar masses.
In contrast to \sersic index, 
\textbf{we make KS tests among all bands, comparing two different bands each time and we found that} 
the $R_{e}$ KS test probabilities are $<$ 0.05 for all morphological types of galaxies (both cluster and non-member galaxies). We conclude that the $R_{e}$ decreases from u to K-band.


Fig. \ref{sample_radius_vulcani_fig16_v1} shows the equivalent diagram as to Fig. \ref{sample_sersic_vulcani_fig9_v1} for $R_{e}$.
The median $\mathcal{R}$ of all morphological types of galaxies, both cluster members and non-members, is smaller in the NIR.
In general, 
$\mathcal{R}$ has values close to 0.7 in the NIR for all galaxies and the largest difference of $\mathcal{R}$ between member and non-member galaxies appears in the K-band but only for ellipticals and late-spirals.
Ellipticals, especially those in clusters, tend to have the largest variation from optical to NIR.  
Lenticulars and early-type spirals present nearly identical 
median
values both inside clusters and in the field 
in all wavelengths.
Non member late-type spirals 
(blue crosses) present the lower variation of $\mathcal{R}$ values while the corresponding cluster members 
(blue asterisks) have larger $\mathcal{R}$ variation than early-type spirals and lenticulars in clusters.
We investigate the influence of stellar mass in Fig. \ref{sample_radius_vulcani_fig16_v1} following the same approach as in Fig. \ref{sample_sersic_vulcani_fig9_v1}.
We applied stellar mass cuts of $log_{10}$($M_{\star}$)=[9,10,11] upon the cluster-member and non-member samples and we find no statistically significant correlation.
In other words, the $R_{e}$ decreases with increasing wavelength independent of the stellar mass.
The only visible difference is the higher value of $\mathcal{R}$ in field galaxies compared to cluster-members for late-type systems. 
However, given the small number of late-type galaxies this should be examined again with larger samples.


\section{Discussion}\label{discussion}

In the previous sections we have presented the multi-wavelength dependence of the morphological structure of cluster galaxies and compared them with non-member galaxies. 
In the next paragraphs, we compare in more detail our findings with previous studies. 

\citet{labarbera10b} studied a sample of bright early-type galaxies ($M_{r}$ < -20 mag) based on multi-wavelength (from g- to K-band) imaging.
They extracted the structural parameters for each band independently and found that the $n$ remains nearly constant with a large scatter (slight increase from optical to NIR) while the $R_{e}$ decreases up to 35\% from g- to K-band. 
Our non-member S0s display a slight increase in the \sersic index (13\%) while the $n$ of non-member ellipticals remains the same in the optical bands and decreases slightly in the K-band.
Regarding $R_{e}$, we also find a decrease as a function of wavelength of $\sim$ 31\% for non-member ellipticals and up to 25\% for non-member S0s. 

\citet{kelvin12} studied the optical and NIR morphology of galaxies in the local Universe extracted from the GAMA survey. 
They fit single-\sersic functions in all bands (from u to K) independently in a magnitude-limited sample ($r \le 19.4$).
For early type galaxies, \citet{kelvin12} find that the $R_{e}$ decreases up to 38\% while we find a change of 31\%$\pm$13 (25\%$\pm$8) decrease in $R_{e}$ for non-member ellipticals (S0s) which is a statistically consistent result.
In addition, \citet{kelvin12} find that the \sersic index of early-type galaxies increases by $\sim$ 52\% while we find that the \sersic index of non-member ellipticals (S0s) stays constant with wavelength: $-6$\%$\pm$10 (13\%$\pm$10).
As for late-type galaxies, they found that the $R_{e}$ decreases up to 25\% from g- to K-band and the \sersic index increases by $\sim$ 52\%.
In our work, the $R_{e}$ decreases by $27\$\pm$5 ($27\$\pm$8) for non-member early spirals (late spirals) statistically consistent result while the \sersic index increases for non-member spiral galaxies (20\%$\pm$15 and 49\%$\pm$33 for early- and late- type spirals respectively).
\citet{vulcani14} applied the same methodology as our study to galaxies with $M_{r}$ < -21.2 and z < 0.3 derived from GAMA survey.
They found that the $R_{e}$ of early-type galaxies decreases up to 45\% which is larger than our result.
On the other hand, \citet{vulcani14} find that the \sersic index of early-type galaxies remains nearly constant as a function of wavelength which is statistically consistent with our result.
However, the $R_{e}$ of late-type systems decreases up to 25\% from u- to H-band (similar to our result) and the \sersic index increases up to 38\% which is close enough to our work.

\citet{kennedy15} studied the wavelength dependence of galaxy structure, using the \citet{vulcani14} derived parameters and focusing on a more local sample (z < 0.15).
They found that the $n$ and $R_{e}$ of the early- type galaxies with $M_r \sim$ -20 decrease (12\% and 23\% respectively) from g- to H-band. 
As they examine brighter galaxies samples ($M_r \sim$-21 and $M_r \sim$-22), the $n$ of early types slight increases (5\%) and the $R_{e}$ decreases (25\%-33\%). 
Moreover, for $M_r \sim$-20 the $n$ of late- type galaxies increases (up to 29\%) while their $R_{e}$ decreases up to 13\%.
The $R_{e}$ of late type galaxies display a similar behaviour with the less bright galaxies (decrease from 15\% to 12\% as a function of wavelenth) while the $n$ increases from 40\% to 55\%.

We summarise the literature results as follows:
In field galaxies there is a clear trend with increasing wavelength. Their size becomes smaller for all galaxy types, while
the \sersic index increases for late-type galaxies and  remains near constant with small variations for the early-type galaxies depending the details of the sample selection.
We note that all the previous studies, that we can compare our results with, have focused on field galaxies and use different proxies in order to define morphology e.g. color and $n$ cuts or visual morphology.
There are, however, very significant difficulties when comparing different studies.
The details governing the sample selection (galaxy mass and morphology) and the exact definition of the environment are important. 
We should keep in mind that subtle differences make direct comparisons challenging and may be responsible for the discrepancies found in different studies.

In our study, we examine two subsets of cluster galaxies. 
The first one include all cluster galaxies of the nine clusters (see Figs. \ref{n_sample_histogram}-\ref{re_vs_LD_multi}) and the second one are the cluster galaxies that have $M_{V}$ <  -19.27 (see Figs. \ref{members_field_sersic_plots}-\ref{sample_radius_vulcani_fig16_v1}).
With the first cluster subsample, we study the structure parameters and the influence of cluster environment. 
The second cluster subsample (bright cluster galaxies) is used in order to compare the values of $n$ and $R_{e}$ with the non-member sample and facilitate the comparison of this study results with previous findings.


The \sersic index of all classes of the first subset of cluster galaxies remains the same as a function of wavelength (optical to NIR). The $R_{e}$ decreases as a function of wavelength. 
The local projected density (even less the $R_{BCG}$) might affect the observed $n$ of cluster member galaxies (S0s and spirals) but not the variation of $n$ with wavelength.
The local projected density and  the $R_{BCG}$ do not affect the $R_{e}$.

The stellar mass plays an important role in the values of structural parameters (as we show in Fig. \ref{sample_n_total_stellar_mass} and Fig. \ref{sample_re_total_stellar_mass}).
More massive galaxies tend to have greater values of \sersic index and $R_{e}$.
The effect is obvious when we compare the first subset of cluster member galaxies  in Fig. \ref{n_sample_multiplots_wt_med}, Fig. \ref{re_sample_multiplots_wt_med}  with the second subset of cluster member galaxies (shown in the asterisks) in Fig. \ref{members_field_sersic_plots} and Fig. \ref{members_field_radius_plots}. 
However, we clearly see that the structural variation with wavelength is statistically the same for each galaxy type that in clusters independent of luminosity (and stellar mass) as also showed in \citet{kennedy15}.

We do not observe any clear trend on structural parameters with projected local density (Fig. \ref{n_vs_LD_multi} and Fig. \ref{re_vs_LD_multi}).
Although small trends exist, such as an increase of the median $n$ for late-type galaxies and a decrease of the median $R_{e}$ for early-type spirals up to log($\Sigma_{10}$) $\sim$ 1.5,  the scatter of these measurements is high. 
Two reasons could be responsible for the lack of clear trends.
Firstly, all galaxies of our cluster member sample are inside $0.64 \times R_{BCG}$, so they are not far away from the cluster center.
An updated analysis of cluster galaxies located in the periphery of clusters (outside $0.64 \times R_{BCG}$) could shed more light on the effect of cluster gravitational potential to the structure of its galaxies.
Secondly, the sample of our nine galaxy clusters is not homogenous in a sense of membership and dynamical relaxation. 
Some clusters have more than 100 members, others have only 30, and only seven of the nine have signs of substructures.

Our analysis of the effects of cluster environment ($R_{BCG}$ and $\Sigma_{10}$) upon the structure reveal that the $n$ and $R_{e}$ of ellipticals and S0s are statistically constant across the $R_{BCG}$ and the local projected density ($\Sigma_{10}$).
As for late spirals, we find minimum values of $n$ and maximum values of $R_{e}$ at the position of $0.3 - 0.4 \times R_{BCG}$. Early spirals have hints for maximum $R_{e}$ values in those regions. The study of \citet{ramella07} showed that seven out of nine galaxy clusters have substructures. We already know that processes like preprocessing and minor merger events are frequent near substructures, influencing the evolution of galaxies. Six out of fourteen substructures of the seven galaxy clusters are located near $0.3 - 0.4 \times R_{BCG}$. Even if the number of late spirals is low, maybe the minimum and maximum values of $n$ at these positions are hints for interesting results about structure parameters near the substructure positions. 
Including more galaxy clusters in an updated study we could study more carefully these results.

In the last part of our analysis, we study the bright cluster galaxies (second sample) comparing them with the non-members galaxies. 
\citet{Park2007} and Davies et. al. (in prep.) found that at fixed morphology and luminosity other physical properties of local galaxies, such as color, color gradient, concentration, size, velocity dispersion, star formation rate and dust content, are nearly independent of local density.
Likewise, \citet{kelkar15}, found no significant difference in the size distribution of cluster and field galaxies of a given morphology.
In this study we find that at fixed morphological bins, bright cluster galaxies show different trends compared to the non-member galaxies.

Bright cluster galaxies have larger \sersic index and smaller $R_{e}$ than their non-member counterparts (see Fig. \ref{members_field_sersic_plots} and Fig. \ref{members_field_radius_plots}).  
This might imply that bright cluster galaxies have more concentrated light profiles compared to non-member galaxies.
In particular, an increase in the \sersic index for spiral galaxies (early and late type) and S0s could be due to the process of adding mass in the central parts of galaxies (gas or stars). 
Tidal interactions (galaxy-galaxy or galaxy-cluster potential) and/or harassment are known to have these effects.

The variation of \sersic index with wavelength is effectively absent for cluster member galaxies in comparison with the variations of non-member galaxies (also in comparison with \citet{vulcani14}.
The absence of \sersic index variation in bright cluster galaxies and the fact that bright cluster galaxies appear more concentrated (higher \sersic indices) compared to non-members counterparts lead us to the following.
If we assume that galaxies located in the field are attracted by the cluster potential and pass by the inner part of the clusters, the effect of environment could change their light distribution.
In particular, material start going to the center of the galaxies \citep{byrdvaltonen90} and the light profiles become more dominant (increase of \sersic index in optical and near-infrared wavelengths) than the light profiles of non-member counterparts. 
At the same time this increase of \sersic index should not happen in the same way both in optical and NIR so we can have a decrease of $\mathcal{N}$ as the galaxy moves from the field into the cluster. 
In addition, another possible scenario is the following: merging of (smaller) galaxies falling into the clusters interact/merge with cluster members increasing their nuclear stellar bulge.

Regarding the sizes, cluster galaxies appear to have small differentiation compared to the non-member galaxies in terms of size variation as a function of wavelength. 
Ellipticals and late-type spirals cluster galaxies are the only exception that have slightly smaller $\mathcal{R}$ compared to non-member galaxies. 
These results show that elliptical galaxies in cluster galaxies have the same $\mathcal{N}$ change as in the field but at the same time have smaller $\mathcal{R}$ values.
This means that the physical process that forces evolution in cluster galaxies changes their size in a different way in optical than in near-infrared while at the same time their light concentration changes at the same way, always compared to the non member galaxies. 
In case of early-type spirals and S0 we see an opposite trend. 
Bright S0s and early-spirals in clusters follow exactly the same size change as a function of wavelength as the non-member equivalent populations (see Fig. 15). 
At the same time these galaxies show an $\mathcal{N}$ slightly higher than 1 in the infrared region which is a value much lower observed in the our non-member galaxies and also in other studies. 
These results are an indication that different morphological types can be affected in a different way by the environment. 
It is remains challenging to assess which processes are responsible for the structural differences between each morphological type of cluster galaxies.

We already know that there is a broad connection between different evolution mechanisms in clusters and the distance from the $R_{BCG}$ \citep{boselligavazzi06}.
Inside the virial radius and close to the center of galaxy clusters (where the density, the temperature of the intergalactic medium (IGM) and the velocity of galaxies have high values), tidal interactions could affect the morphology of galaxies and consequently their light-profiles.
Galaxy-galaxy interactions, even if they take place near the center of rich clusters, remain fairly rare ($t_{enc} \sim 10^{8} yr$) and their relative velocities of the encounters are large compared to their typical rotational speeds. 
The most efficient processes close to the center of a cluster is the galaxy-cluster IGM interactions, harassment and ram-pressure.
Our findings support the idea of the above processes for cluster galaxies (spirals and S0s) if we compare the $n$ and $R_{e}$ with their non-member counterparts. 

An upcoming study on the OMEGACAM data (largee FoV $1 \sim degree$) will increase the number of WINGS galaxies and it will shed more light in the difference between local cluster galaxies and the interpretation of the physical mechanisms that affect the evolution of galaxies in the peripheral regions.

\section{Conclusions}\label{conclusions}

In this paper we study the structural parameters of cluster galaxies across optical and NIR wavelengths. 
The structural parameters have been calculated with the state-of -the-art software \galapagosII by using single-\sersic functions (from u- to K-band).

\indent Our multi-wavelength analysis of nine galaxy clusters shows that:
\begin{enumerate}
\item 
The weighted median value of \sersic index of cluster member galaxies stays constant across optical and NIR wavelengths ( see upper panel of Fig. \ref{n_sample_multiplots_wt_med})
while the median values of $R_{e}$ tend to decrease for all morphological types (see upper panel of Fig. \ref{re_sample_multiplots_wt_med}).

\item 
Late-type spirals in the inner region of the clusters as well as in the higher density regions show a slight increase in the average \sersic index compared to outer and lower density regions of the clusters respectively.
\sersic index of lenticulars and early-type spirals seems to be affected only by the environment density.
Lenticulars increase their average \sersic index in the higher density environment, like late-type spirals do, while early-type spirals show the opposite trend (see middle and down panels of Fig.\ref{n_sample_multiplots_wt_med}).

\item 
Late- type spiral galaxies appear to have smaller average $R_{e}$ when they are at the outer area of the cluster, even though the difference is within the error bars.
The median $R_{e}$ is larger in higher local densities rather than in regions with lower local densities.
In contrast, the size of ellipticals, S0s, and early-type spirals do not change noticeable as a function of the environment (see middle and down panels of Fig.\ref{re_sample_multiplots_wt_med}).

\item 
Late-type spirals display a decrease in their median \sersic index as a function with the distance from the cluster center up to $\sim 0.3\times R_{BCG}$ in the optical bands, while further out $n$ remains statistically constant (see Fig.\ref{n_vs_rBCG_multi}).
Early-type spirals have constant $R_{e}$ values across $R_{BCG}$ while late-type spirals increase their median $R_{e}$ up to $\sim 0.35 \times R_{BCG}$ in optical and especially in NIR bands (see Fig.\ref{re_vs_rBCG_multi}).

\item 
We do not observe any clear trend of structural parameters with projected local density (see Fig. \ref{n_vs_LD_multi} and Fig. \ref{re_vs_LD_multi}).

\item 
Early and late type spirals, as well as S0s, which are cluster members fulfilling the $M_{V}$=-19.27mag selection limit, display on average greater \sersic index (see Fig. \ref{members_field_sersic_plots}) and smaller $R_{e}$ than their non-member counterparts (see Fig. \ref{members_field_radius_plots}).
This is not the case for ellipticals. 

\item 

Despite the apparent trends seen in Fig. \ref{members_field_sersic_plots} and Fig. \ref{members_field_radius_plots}, no statistically significant changes are observed in the variation of the \sersic index and $R_{e}$ for any galaxy type fulfilling the $M_{V}$=-19.27 mag selection limit, as a function of wavelength.

\end{enumerate}

\begin{acknowledgements}
The authors are grateful to the referee, whose detailed and constructive comments greatly improved the manuscript. 
We would also like to thank T. Bitsakis for many useful discussions and comments on this work.
This research made use of \montage. It is funded by the National Science Foundation under Grant Number ACI-1440620, and was previously funded by the National Aeronautics and Space Administration's Earth Science Technology Office, Computation Technologies Project, under Cooperative Agreement Number NCC5-626 between NASA and the California Institute of Technology. M.V. was supported by an IKY postdoctoral Scholarship.
\end{acknowledgements}


\bibliographystyle{aa}
\bibliography{psychogyios_et_al_wings_v25}

\begin{thebibliography}{66}
\expandafter\ifx\csname natexlab\endcsname\relax\def\natexlab#1{#1}\fi

\bibitem[{{Bamford} {et~al.}(2012){Bamford}, {H{\"a}u{\ss}ler}, {Rojas},
  {Vika}, \& {Cresswell}}]{bamford12}
{Bamford}, S.~P., {H{\"a}u{\ss}ler}, B., {Rojas}, A., {Vika}, M., \&
  {Cresswell}, J. 2012, in IAU Symposium, Vol. 284, The Spectral Energy
  Distribution of Galaxies - SED 2011, ed. R.~J. {Tuffs} \& C.~C. {Popescu},
  301--305

\bibitem[{{Barden} {et~al.}(2012){Barden}, {H{\"a}u{\ss}ler}, {Peng},
  {McIntosh}, \& {Guo}}]{barden12}
{Barden}, M., {H{\"a}u{\ss}ler}, B., {Peng}, C.~Y., {McIntosh}, D.~H., \&
  {Guo}, Y. 2012, \mnras, 422, 449

\bibitem[{{Bekki}(1998)}]{bekki98}
{Bekki}, K. 1998, \apjl, 502, L133

\bibitem[{{Bell} {et~al.}(2003){Bell}, {McIntosh}, {Katz}, \&
  {Weinberg}}]{bell03}
{Bell}, E.~F., {McIntosh}, D.~H., {Katz}, N., \& {Weinberg}, M.~D. 2003, \apjs,
  149, 289

\bibitem[{{Berriman} {et~al.}(2008){Berriman}, {Good}, {Laity}, \&
  {Kong}}]{berriman08}
{Berriman}, G.~B., {Good}, J.~C., {Laity}, A.~C., \& {Kong}, M. 2008, in
  Astronomical Society of the Pacific Conference Series, Vol. 394, Astronomical
  Data Analysis Software and Systems XVII, ed. R.~W. {Argyle}, P.~S.
  {Bunclark}, \& J.~R. {Lewis}, 83

\bibitem[{{Bertin}(2011)}]{bertin11}
{Bertin}, E. 2011, in Astronomical Society of the Pacific Conference Series,
  Vol. 442, Astronomical Data Analysis Software and Systems XX, ed. I.~N.
  {Evans}, A.~{Accomazzi}, D.~J. {Mink}, \& A.~H. {Rots}, 435

\bibitem[{{Bertin} \& {Arnouts}(1996)}]{bertin96}
{Bertin}, E. \& {Arnouts}, S. 1996, \aaps, 117, 393

\bibitem[{{Boselli} \& {Gavazzi}(2006)}]{boselligavazzi06}
{Boselli}, A. \& {Gavazzi}, G. 2006, \pasp, 118, 517

\bibitem[{{Butcher} \& {Oemler}(1978)}]{butcher_oemler}
{Butcher}, H. \& {Oemler}, Jr., A. 1978, \apj, 219, 18

\bibitem[{{Byrd} \& {Valtonen}(1990)}]{byrdvaltonen90}
{Byrd}, G. \& {Valtonen}, M. 1990, \apj, 350, 89

\bibitem[{{Calvi} {et~al.}(2013){Calvi}, {Poggianti}, {Vulcani}, \&
  {Fasano}}]{calvi13}
{Calvi}, R., {Poggianti}, B.~M., {Vulcani}, B., \& {Fasano}, G. 2013, \mnras,
  432, 3141

\bibitem[{{Cava} {et~al.}(2009){Cava}, {Bettoni}, {Poggianti}, {Couch},
  {Moles}, {Varela}, {Biviano}, {D'Onofrio}, {Dressler}, {Fasano}, {Fritz},
  {Kj{\ae}rgaard}, {Ramella}, \& {Valentinuzzi}}]{cava09}
{Cava}, A., {Bettoni}, D., {Poggianti}, B.~M., {et~al.} 2009, \aap, 495, 707

\bibitem[{{Cerulo} {et~al.}(2017){Cerulo}, {Couch}, {Lidman}, {Demarco},
  {Huertas-Company}, {Mei}, {S{\'a}nchez-Janssen}, {Barrientos}, \&
  {Mu{\~n}oz}}]{cerulo17}
{Cerulo}, P., {Couch}, W.~J., {Lidman}, C., {et~al.} 2017, ArXiv e-prints
  [\eprint[arXiv]{1707.00751}]

\bibitem[{{Desai} {et~al.}(2007){Desai}, {Dalcanton}, {Arag{\'o}n-Salamanca},
  {Jablonka}, {Poggianti}, {Gogarten}, {Simard}, {Milvang-Jensen}, {Rudnick},
  {Zaritsky}, {Clowe}, {Halliday}, {Pell{\'o}}, {Saglia}, \& {White}}]{desai07}
{Desai}, V., {Dalcanton}, J.~J., {Arag{\'o}n-Salamanca}, A., {et~al.} 2007,
  \apj, 660, 1151

\bibitem[{{D'Onofrio} {et~al.}(2014){D'Onofrio}, {Bindoni}, {Fasano},
  {Bettoni}, {Cava}, {Fritz}, {Gullieuszik}, {Kj{\ae}rgaard}, {Moretti},
  {Moles}, {Omizzolo}, {Poggianti}, {Valentinuzzi}, \& {Varela}}]{donofrio14}
{D'Onofrio}, M., {Bindoni}, D., {Fasano}, G., {et~al.} 2014, \aap, 572, A87

\bibitem[{{Dressler}(1980)}]{dressler80}
{Dressler}, A. 1980, \apjs, 42, 565

\bibitem[{{Dressler} {et~al.}(1997){Dressler}, {Oemler}, {Couch}, {Smail},
  {Ellis}, {Barger}, {Butcher}, {Poggianti}, \& {Sharples}}]{dressler97}
{Dressler}, A., {Oemler}, Jr., A., {Couch}, W.~J., {et~al.} 1997, \apj, 490,
  577

\bibitem[{{Driver} {et~al.}(2009){Driver}, {Norberg}, {Baldry}, {Bamford},
  {Hopkins}, {Liske}, {Loveday}, {Peacock}, {Hill}, {Kelvin}, {Robotham},
  {Cross}, {Parkinson}, {Prescott}, {Conselice}, {Dunne}, {Brough}, {Jones},
  {Sharp}, {van Kampen}, {Oliver}, {Roseboom}, {Bland-Hawthorn}, {Croom},
  {Ellis}, {Cameron}, {Cole}, {Frenk}, {Couch}, {Graham}, {Proctor}, {De
  Propris}, {Doyle}, {Edmondson}, {Nichol}, {Thomas}, {Eales}, {Jarvis},
  {Kuijken}, {Lahav}, {Madore}, {Seibert}, {Meyer}, {Staveley-Smith},
  {Phillipps}, {Popescu}, {Sansom}, {Sutherland}, {Tuffs}, \&
  {Warren}}]{driver09}
{Driver}, S.~P., {Norberg}, P., {Baldry}, I.~K., {et~al.} 2009, Astronomy and
  Geophysics, 50, 5.12

\bibitem[{{Fasano} {et~al.}(2006){Fasano}, {Marmo}, {Varela}, {D'Onofrio},
  {Poggianti}, {Moles}, {Pignatelli}, {Bettoni}, {Kj{\ae}rgaard}, {Rizzi},
  {Couch}, \& {Dressler}}]{fasano06}
{Fasano}, G., {Marmo}, C., {Varela}, J., {et~al.} 2006, \aap, 445, 805

\bibitem[{{Fasano} {et~al.}(2015){Fasano}, {Poggianti}, {Bettoni}, {D'Onofrio},
  {Dressler}, {Vulcani}, {Moretti}, {Gullieuszik}, {Fritz}, {Omizzolo}, {Cava},
  {Couch}, {Ramella}, \& {Biviano}}]{fasano15}
{Fasano}, G., {Poggianti}, B.~M., {Bettoni}, D., {et~al.} 2015, \mnras, 449,
  3927

\bibitem[{{Fasano} {et~al.}(2000){Fasano}, {Poggianti}, {Couch}, {Bettoni},
  {Kj{\ae}rgaard}, \& {Moles}}]{fasano00}
{Fasano}, G., {Poggianti}, B.~M., {Couch}, W.~J., {et~al.} 2000, \apj, 542, 673

\bibitem[{{Fasano} {et~al.}(2012){Fasano}, {Vanzella}, {Dressler}, {Poggianti},
  {Moles}, {Bettoni}, {Valentinuzzi}, {Moretti}, {D'Onofrio}, {Varela},
  {Couch}, {Kj{\ae}rgaard}, {Fritz}, {Omizzolo}, \& {Cava}}]{fasano12}
{Fasano}, G., {Vanzella}, E., {Dressler}, A., {et~al.} 2012, \mnras, 420, 926

\bibitem[{{Fritz} {et~al.}(2007){Fritz}, {Poggianti}, {Bettoni}, {Cava},
  {Couch}, {D'Onofrio}, {Dressler}, {Fasano}, {Kj{\ae}rgaard}, {Moles}, \&
  {Varela}}]{fritz07}
{Fritz}, J., {Poggianti}, B.~M., {Bettoni}, D., {et~al.} 2007, \aap, 470, 137

\bibitem[{{Fritz} {et~al.}(2014){Fritz}, {Poggianti}, {Cava}, {Moretti},
  {Varela}, {Bettoni}, {Couch}, {D'Onofrio D'Onofrio}, {Dressler}, {Fasano},
  {Kj{\ae}rgaard}, {Marziani}, {Moles}, \& {Omizzolo}}]{fritz14}
{Fritz}, J., {Poggianti}, B.~M., {Cava}, A., {et~al.} 2014, \aap, 566, A32

\bibitem[{{Fritz} {et~al.}(2011){Fritz}, {Poggianti}, {Cava}, {Valentinuzzi},
  {Moretti}, {Bettoni}, {Bressan}, {Couch}, {D'Onofrio}, {Dressler}, {Fasano},
  {Kj{\ae}rgaard}, {Moles}, {Omizzolo}, \& {Varela}}]{fritz11}
{Fritz}, J., {Poggianti}, B.~M., {Cava}, A., {et~al.} 2011, \aap, 526, A45

\bibitem[{{Gunn} \& {Gott}(1972)}]{gunn_gott_72}
{Gunn}, J.~E. \& {Gott}, III, J.~R. 1972, \apj, 176, 1

\bibitem[{{H{\"a}u{\ss}ler} {et~al.}(2013){H{\"a}u{\ss}ler}, {Bamford}, {Vika},
  {Rojas}, {Barden}, {Kelvin}, {Alpaslan}, {Robotham}, {Driver}, {Baldry},
  {Brough}, {Hopkins}, {Liske}, {Nichol}, {Popescu}, \& {Tuffs}}]{Haussler13}
{H{\"a}u{\ss}ler}, B., {Bamford}, S.~P., {Vika}, M., {et~al.} 2013, \mnras,
  430, 330

\bibitem[{{H{\"a}ussler} {et~al.}(2007){H{\"a}ussler}, {McIntosh}, {Barden},
  {Bell}, {Rix}, {Borch}, {Beckwith}, {Caldwell}, {Heymans}, {Jahnke}, {Jogee},
  {Koposov}, {Meisenheimer}, {S{\'a}nchez}, {Somerville}, {Wisotzki}, \&
  {Wolf}}]{haussler07}
{H{\"a}ussler}, B., {McIntosh}, D.~H., {Barden}, M., {et~al.} 2007, \apjs, 172,
  615

\bibitem[{{Hubble} \& {Humason}(1931)}]{hubble_humason_31}
{Hubble}, E. \& {Humason}, M.~L. 1931, \apj, 74, 43

\bibitem[{{Icke}(1985)}]{icke85}
{Icke}, V. 1985, \aap, 144, 115

\bibitem[{{Kelkar} {et~al.}(2015){Kelkar}, {Arag{\'o}n-Salamanca}, {Gray},
  {Maltby}, {Vulcani}, {De Lucia}, {Poggianti}, \& {Zaritsky}}]{kelkar15}
{Kelkar}, K., {Arag{\'o}n-Salamanca}, A., {Gray}, M.~E., {et~al.} 2015, \mnras,
  450, 1246

\bibitem[{{Kelvin} {et~al.}(2012){Kelvin}, {Driver}, {Robotham}, {Hill},
  {Alpaslan}, {Baldry}, {Bamford}, {Bland-Hawthorn}, {Brough}, {Graham},
  {H{\"a}ussler}, {Hopkins}, {Liske}, {Loveday}, {Norberg}, {Phillipps},
  {Popescu}, {Prescott}, {Taylor}, \& {Tuffs}}]{kelvin12}
{Kelvin}, L.~S., {Driver}, S.~P., {Robotham}, A.~S.~G., {et~al.} 2012, \mnras,
  421, 1007

\bibitem[{{Kennedy} {et~al.}(2015){Kennedy}, {Bamford}, {Baldry},
  {H{\"a}u{\ss}ler}, {Holwerda}, {Hopkins}, {Kelvin}, {Lange}, {Moffett},
  {Popescu}, {Taylor}, {Tuffs}, {Vika}, \& {Vulcani}}]{kennedy15}
{Kennedy}, R., {Bamford}, S.~P., {Baldry}, I., {et~al.} 2015, \mnras, 454, 806

\bibitem[{{Kim} \& {Im}(2013)}]{Kim2013}
{Kim}, D. \& {Im}, M. 2013, \apj, 766, 109

\bibitem[{{La Barbera} {et~al.}(2010){La Barbera}, {de Carvalho}, {de La Rosa},
  {Lopes}, {Kohl-Moreira}, \& {Capelato}}]{labarbera10b}
{La Barbera}, F., {de Carvalho}, R.~R., {de La Rosa}, I.~G., {et~al.} 2010,
  \mnras, 408, 1313

\bibitem[{{Larson} {et~al.}(1980){Larson}, {Tinsley}, \& {Caldwell}}]{larson80}
{Larson}, R.~B., {Tinsley}, B.~M., \& {Caldwell}, C.~N. 1980, \apj, 237, 692

\bibitem[{{Lubin} {et~al.}(2002){Lubin}, {Oke}, \& {Postman}}]{lubin02}
{Lubin}, L.~M., {Oke}, J.~B., \& {Postman}, M. 2002, \aj, 124, 1905

\bibitem[{{MacArthur} {et~al.}(2004){MacArthur}, {Courteau}, {Bell}, \&
  {Holtzman}}]{MacArthur04}
{MacArthur}, L.~A., {Courteau}, S., {Bell}, E., \& {Holtzman}, J.~A. 2004,
  \apjs, 152, 175

\bibitem[{{Moore} {et~al.}(1996){Moore}, {Katz}, {Lake}, {Dressler}, \&
  {Oemler}}]{moore96}
{Moore}, B., {Katz}, N., {Lake}, G., {Dressler}, A., \& {Oemler}, A. 1996,
  \nat, 379, 613

\bibitem[{{Moore} {et~al.}(1998){Moore}, {Lake}, \& {Katz}}]{moore98}
{Moore}, B., {Lake}, G., \& {Katz}, N. 1998, \apj, 495, 139

\bibitem[{{Moore} {et~al.}(1999){Moore}, {Lake}, {Quinn}, \&
  {Stadel}}]{moore99}
{Moore}, B., {Lake}, G., {Quinn}, T., \& {Stadel}, J. 1999, \mnras, 304, 465

\bibitem[{{Moretti} {et~al.}(2017){Moretti}, {Gullieuszik}, {Poggianti},
  {Paccagnella}, {Couch}, {Vulcani}, {Bettoni}, {Fritz}, {Cava}, {Fasano},
  {D'Onofrio}, \& {Omizzolo}}]{moretti17}
{Moretti}, A., {Gullieuszik}, M., {Poggianti}, B., {et~al.} 2017, \aap, 599,
  A81

\bibitem[{{Moretti} {et~al.}(2014){Moretti}, {Poggianti}, {Fasano}, {Bettoni},
  {D'Onofrio}, {Fritz}, {Cava}, {Varela}, {Vulcani}, {Gullieuszik}, {Couch},
  {Omizzolo}, {Valentinuzzi}, {Dressler}, {Moles}, {Kj{\ae}rgaard},
  {Smareglia}, \& {Molinaro}}]{moretti14}
{Moretti}, A., {Poggianti}, B.~M., {Fasano}, G., {et~al.} 2014, \aap, 564, A138

\bibitem[{{Omizzolo} {et~al.}(2014){Omizzolo}, {Fasano}, {Reverte Paya}, {De
  Santis}, {Grado}, {Bettoni}, {Poggianti}, {D'Onofrio}, {Moretti}, {Varela},
  {Fritz}, {Gullieuszik}, {Cava}, {Grazian}, \& {Moles}}]{omizzolo14}
{Omizzolo}, A., {Fasano}, G., {Reverte Paya}, D., {et~al.} 2014, \aap, 561,
  A111

\bibitem[{{Park} {et~al.}(2007){Park}, {Choi}, {Vogeley}, {Gott}, {Blanton}, \&
  {SDSS Collaboration}}]{Park2007}
{Park}, C., {Choi}, Y.-Y., {Vogeley}, M.~S., {et~al.} 2007, \apj, 658, 898

\bibitem[{{Peng} {et~al.}(2002){Peng}, {Ho}, {Impey}, \& {Rix}}]{peng02}
{Peng}, C.~Y., {Ho}, L.~C., {Impey}, C.~D., \& {Rix}, H.-W. 2002, \aj, 124, 266

\bibitem[{{Pignatelli} {et~al.}(2006){Pignatelli}, {Fasano}, \&
  {Cassata}}]{pignatelli06}
{Pignatelli}, E., {Fasano}, G., \& {Cassata}, P. 2006, \aap, 446, 373

\bibitem[{{Poggianti} {et~al.}(2009){Poggianti}, {Fasano}, {Bettoni}, {Cava},
  {Dressler}, {Vanzella}, {Varela}, {Couch}, {D'Onofrio}, {Fritz},
  {Kjaergaard}, {Moles}, \& {Valentinuzzi}}]{poggianti09}
{Poggianti}, B.~M., {Fasano}, G., {Bettoni}, D., {et~al.} 2009, \apjl, 697,
  L137

\bibitem[{{Postman} {et~al.}(2005){Postman}, {Franx}, {Cross}, {Holden},
  {Ford}, {Illingworth}, {Goto}, {Demarco}, {Rosati}, {Blakeslee}, {Tran},
  {Ben{\'{\i}}tez}, {Clampin}, {Hartig}, {Homeier}, {Ardila}, {Bartko},
  {Bouwens}, {Bradley}, {Broadhurst}, {Brown}, {Burrows}, {Cheng}, {Feldman},
  {Golimowski}, {Gronwall}, {Infante}, {Kimble}, {Krist}, {Lesser}, {Martel},
  {Mei}, {Menanteau}, {Meurer}, {Miley}, {Motta}, {Sirianni}, {Sparks}, {Tran},
  {Tsvetanov}, {White}, \& {Zheng}}]{postman05}
{Postman}, M., {Franx}, M., {Cross}, N.~J.~G., {et~al.} 2005, \apj, 623, 721

\bibitem[{{Ramella} {et~al.}(2007){Ramella}, {Biviano}, {Pisani}, {Varela},
  {Bettoni}, {Couch}, {D'Onofrio}, {Dressler}, {Fasano}, {Kj{\o}rgaard},
  {Moles}, {Pignatelli}, \& {Poggianti}}]{ramella07}
{Ramella}, M., {Biviano}, A., {Pisani}, A., {et~al.} 2007, \aap, 470, 39

\bibitem[{{Robotham} {et~al.}(2011){Robotham}, {Norberg}, {Driver}, {Baldry},
  {Bamford}, {Hopkins}, {Liske}, {Loveday}, {Merson}, {Peacock}, {Brough},
  {Cameron}, {Conselice}, {Croom}, {Frenk}, {Gunawardhana}, {Hill}, {Jones},
  {Kelvin}, {Kuijken}, {Nichol}, {Parkinson}, {Pimbblet}, {Phillipps},
  {Popescu}, {Prescott}, {Sharp}, {Sutherland}, {Taylor}, {Thomas}, {Tuffs},
  {van Kampen}, \& {Wijesinghe}}]{robotham11}
{Robotham}, A.~S.~G., {Norberg}, P., {Driver}, S.~P., {et~al.} 2011, \mnras,
  416, 2640

\bibitem[{{Schmidt} {et~al.}(1997){Schmidt}, {Bohm}, \& {Elsasser}}]{schmidt97}
{Schmidt}, K.-H., {Bohm}, P., \& {Elsasser}, H. 1997, Astronomische
  Nachrichten, 318, 81

\bibitem[{{Sersic} {et~al.}(1968){Sersic}, {Pastoriza}, \&
  {Carranza}}]{Sersic68}
{Sersic}, J.~L., {Pastoriza}, M.~G., \& {Carranza}, G.~J. 1968, \aplett, 2, 45

\bibitem[{{Smail} {et~al.}(1997){Smail}, {Dressler}, {Couch}, {Ellis},
  {Oemler}, {Butcher}, \& {Sharples}}]{smail97}
{Smail}, I., {Dressler}, A., {Couch}, W.~J., {et~al.} 1997, \apjs, 110, 213

\bibitem[{{Valentinuzzi} {et~al.}(2009){Valentinuzzi}, {Woods}, {Fasano},
  {Riello}, {D'Onofrio}, {Varela}, {Bettoni}, {Cava}, {Couch}, {Dressler},
  {Fritz}, {Moles}, {Omizzolo}, {Poggianti}, \&
  {Kj{\ae}rgaard}}]{valentinuzzi09}
{Valentinuzzi}, T., {Woods}, D., {Fasano}, G., {et~al.} 2009, \aap, 501, 851

\bibitem[{{van Dokkum} {et~al.}(2000){van Dokkum}, {Franx}, {Fabricant},
  {Illingworth}, \& {Kelson}}]{vandokkum00}
{van Dokkum}, P.~G., {Franx}, M., {Fabricant}, D., {Illingworth}, G.~D., \&
  {Kelson}, D.~D. 2000, \apj, 541, 95

\bibitem[{{Varela} {et~al.}(2009){Varela}, {D'Onofrio}, {Marmo}, {Fasano},
  {Bettoni}, {Cava}, {Couch}, {Dressler}, {Kj{\ae}rgaard}, {Moles},
  {Pignatelli}, {Poggianti}, \& {Valentinuzzi}}]{varela09}
{Varela}, J., {D'Onofrio}, M., {Marmo}, C., {et~al.} 2009, \aap, 497, 667

\bibitem[{{Vika} {et~al.}(2014){Vika}, {Bamford}, {H{\"a}u{\ss}ler}, \&
  {Rojas}}]{vika14}
{Vika}, M., {Bamford}, S.~P., {H{\"a}u{\ss}ler}, B., \& {Rojas}, A.~L. 2014,
  \mnras, 444, 3603

\bibitem[{{Vika} {et~al.}(2013){Vika}, {Bamford}, {H{\"a}u{\ss}ler}, {Rojas},
  {Borch}, \& {Nichol}}]{vika13}
{Vika}, M., {Bamford}, S.~P., {H{\"a}u{\ss}ler}, B., {et~al.} 2013, \mnras,
  435, 623

\bibitem[{{Vika} {et~al.}(2015){Vika}, {Vulcani}, {Bamford}, {H{\"a}u{\ss}ler},
  \& {Rojas}}]{vika15}
{Vika}, M., {Vulcani}, B., {Bamford}, S.~P., {H{\"a}u{\ss}ler}, B., \& {Rojas},
  A.~L. 2015, \aap, 577, A97

\bibitem[{{Vulcani} {et~al.}(2014){Vulcani}, {Bamford}, {H{\"a}u{\ss}ler},
  {Vika}, {Rojas}, {Agius}, {Baldry}, {Bauer}, {Brown}, {Driver}, {Graham},
  {Kelvin}, {Liske}, {Loveday}, {Popescu}, {Robotham}, \& {Tuffs}}]{vulcani14}
{Vulcani}, B., {Bamford}, S.~P., {H{\"a}u{\ss}ler}, B., {et~al.} 2014, \mnras,
  441, 1340

\bibitem[{{Vulcani} {et~al.}(2011{\natexlab{a}}){Vulcani}, {Poggianti},
  {Arag{\'o}n-Salamanca}, {Fasano}, {Rudnick}, {Valentinuzzi}, {Dressler},
  {Bettoni}, {Cava}, {D'Onofrio}, {Fritz}, {Moretti}, {Omizzolo}, \&
  {Varela}}]{vulcani11a}
{Vulcani}, B., {Poggianti}, B.~M., {Arag{\'o}n-Salamanca}, A., {et~al.}
  2011{\natexlab{a}}, \mnras, 412, 246

\bibitem[{{Vulcani} {et~al.}(2011{\natexlab{b}}){Vulcani}, {Poggianti},
  {Dressler}, {Fasano}, {Valentinuzzi}, {Couch}, {Moretti}, {Simard}, {Desai},
  {Bettoni}, {D'Onofrio}, {Cava}, \& {Varela}}]{vulcani11b}
{Vulcani}, B., {Poggianti}, B.~M., {Dressler}, A., {et~al.} 2011{\natexlab{b}},
  \mnras, 413, 921

\bibitem[{{Vulcani} {et~al.}(2012){Vulcani}, {Poggianti}, {Fasano}, {Desai},
  {Dressler}, {Oemler}, {Calvi}, {D'Onofrio}, \& {Moretti}}]{vulcani12a}
{Vulcani}, B., {Poggianti}, B.~M., {Fasano}, G., {et~al.} 2012, \mnras, 420,
  1481

\bibitem[{{Vulcani} {et~al.}(2013){Vulcani}, {Poggianti}, {Oemler}, {Dressler},
  {Arag{\'o}n-Salamanca}, {De Lucia}, {Moretti}, {Gladders}, {Abramson}, \&
  {Halliday}}]{vulcani13}
{Vulcani}, B., {Poggianti}, B.~M., {Oemler}, A., {et~al.} 2013, \aap, 550, A58

\bibitem[{{Whitmore} {et~al.}(1993){Whitmore}, {Gilmore}, \&
  {Jones}}]{whitmore93}
{Whitmore}, B.~C., {Gilmore}, D.~M., \& {Jones}, C. 1993, \apj, 407, 489

\end{thebibliography}

\begin{appendix} 
\section{\textbf{Comparison of} \galapagosII with GASPHOT}
\label{appa}

\citet{donofrio14} applied the software called GAlaxy Surface PHOTometry \citep[GASPHOT;][]{pignatelli06}, on B-, V- and K-band images of WINGS galaxies. Here we compare the structural parameters of \galapagosII with GASPHOT. Fig. \ref{gala_gas_v} shows the magnitude, $n$ and $R_{e}$ of the cross-matced cluster galaxies in V-band.


\begin{figure}
{\includegraphics[width=\hsize]{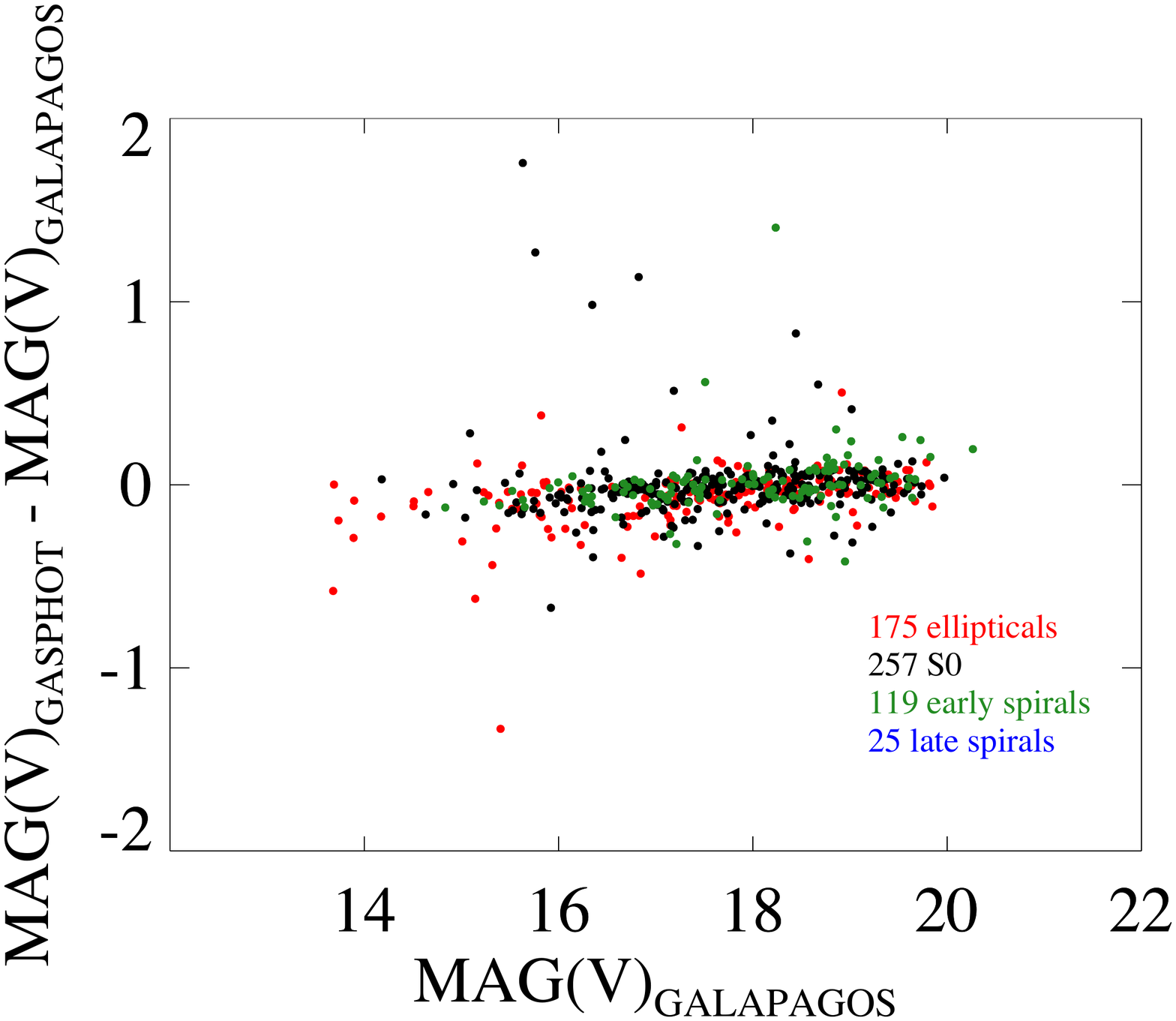}}        
{\includegraphics[width=\hsize]{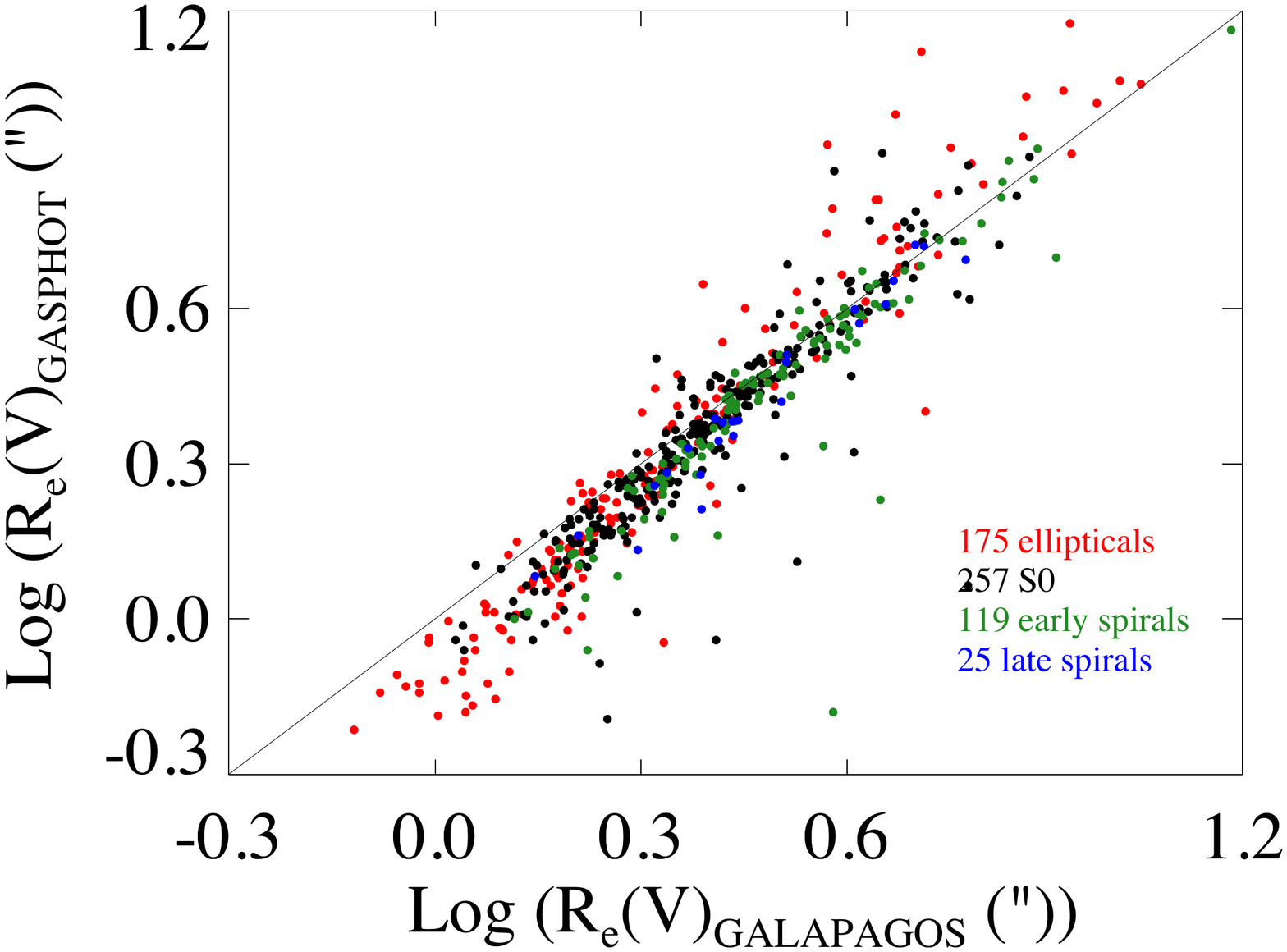}}
{\includegraphics[width=\hsize]{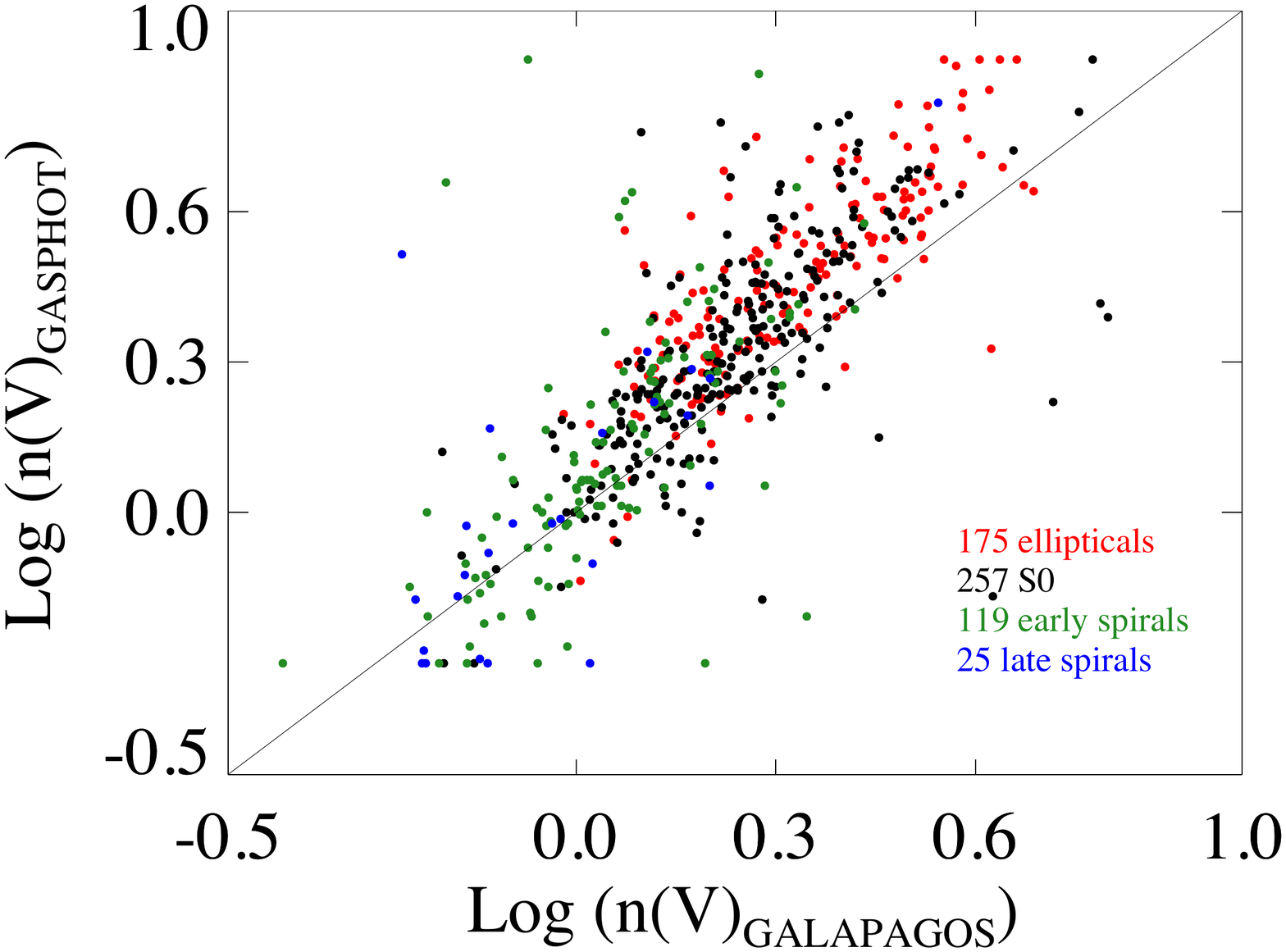}}
\caption{Structural parameters for all galaxies in V-band.}
\label{gala_gas_v}
\end{figure}

GASPHOT and \galfitm measure the same magnitudes for all galaxy types.
GASPHOT measures smaller sizes for small elliptical galaxies.
In addition, we see that for galaxies with $n$ > 2, GASPHOT measures larger light profiles compare to \galfitm. 
The above discrepancy could be due to various effects \citep{pignatelli06,haussler07}. 
The most important is GASPHOT uses the 1D approach while \galfitm utilises a 2D method. 
Second, the sky has been measured in two different ways, which can lead to substantial differences in the fit parameters \citep{haussler07}. 
Finally, we must not forget that the both softwares have average uncertainties of the structural parameters as \citet{donofrio14} showed. The uncerainty range for the $R_{e}$ is from 7\% to 20\% while for the $n$ is from 20\% to 30\%. In addition, \citet{donofrio14} showed that these two methods have a different sensitivity to the peculiar features of galaxies and behave differently in weighting the various (inner and outer) galaxy regions.

\end{appendix}


\end{document}